\documentclass[journal]{IEEEtran}
\usepackage{graphicx,dblfloatfix}
\usepackage{float}
\usepackage{amsmath}
\usepackage[linesnumbered,ruled]{algorithm2e}
\usepackage[sort]{cite} 
\nocite{*} 

\ifCLASSINFOpdf
\else
\fi
\usepackage{graphicx} 
\usepackage{booktabs} %
\usepackage{multirow}
\usepackage{textcomp}
\usepackage[linesnumbered,ruled]{algorithm2e}
\begin{document}
%
\title{Spectral Zooming and Resolution Limits of Spatial Spectral Compressive Spectral Imagers}
\author{Edgar~Salazar,
	Alejandro Parada-Mayorga
	and~Gonzalo R. Arce}


%


\maketitle

\begin{abstract}	
The recently introduced \textit{Spatial Spectral Compressive Spectral
Imager (SSCSI)}  has been proposed as an alternative to carry out spatial and spectral coding using a binary on-off coded aperture. In SSCSI, the pixel pitch size of the coded aperture, as well as its location 
with respect to the detector array, play a critical role in the quality of image reconstruction. In this paper, 
a rigorous discretization
model for this architecture is developed, based on a light propagation analysis across the imager. The attainable spatial and 
spectral resolution, and the various parameters affecting them, is derived through this process. Much like the displacement of zoom lens components
leads to higher spatial resolution of a scene, a shift of the coded aperture in the SSCSI in reference to the detector
leads to higher spectral resolution. This allows the recovery of spectrally detailed datacubes by physically displacing the mask towards the spectral plane.
To prove the underlying concepts, computer simulations and experimental data are presented in this paper.
\end{abstract}


%
\IEEEpeerreviewmaketitle

\section{Introduction}
 Hyperspectral imaging allows the simultaneous acquisition of spatial and spectral information of a given scene. It finds applications  in a variety of fields such as remote sensing, medical imaging, food quality and spectroscopy \cite{shaw2003spectral,lorente,lu2014medical}. As an alternative to the conventional hyperspectral sensing methods (push-broom, whisk-broom),  the \textit{Coded Aperture Snapshot Spectral Imager} (CASSI)  \cite{brady,bradyoptical,thesis,arce_intro} was introduced; this is a Compressive Sensing based architecture \cite{donohocompressed}, where the entire spatial-spectral datacube can be recovered from a few sensor measurements. A recently proposed enhancement of CASSI, replaces the
 binary on-off coded aperture by a pixelated color filter mask to achieve three dimensional coding prior to the optical sensing. This imager, known as colored CASSI \cite{colored,rueda2015dmd,aparada,rueda2017high}, increases both the
 quality of the reconstruction in the spatial domain and the 
 attainable spectral resolution. Additionally, fewer snapshots are required to obtain accurate results \cite{colored}. The \textit{Snapshot Colored Compressive Spectral Imager} (SCCSI)\cite{correa2015} and \cite{correa_multiple}, on the other hand, represents a more compact version of the colored CASSI with remarkable performance, where the coded aperture is directly located on top of the sensor. Despite the advantages of both the colored CASSI and the SCCSI, the high fabrication 
 cost of pixelated
 colored coded apertures limits their implementation in practice\cite{BARRIE19956}. 
 
\begin{figure}[h!]
 	\centering
 	\includegraphics[scale=0.16]{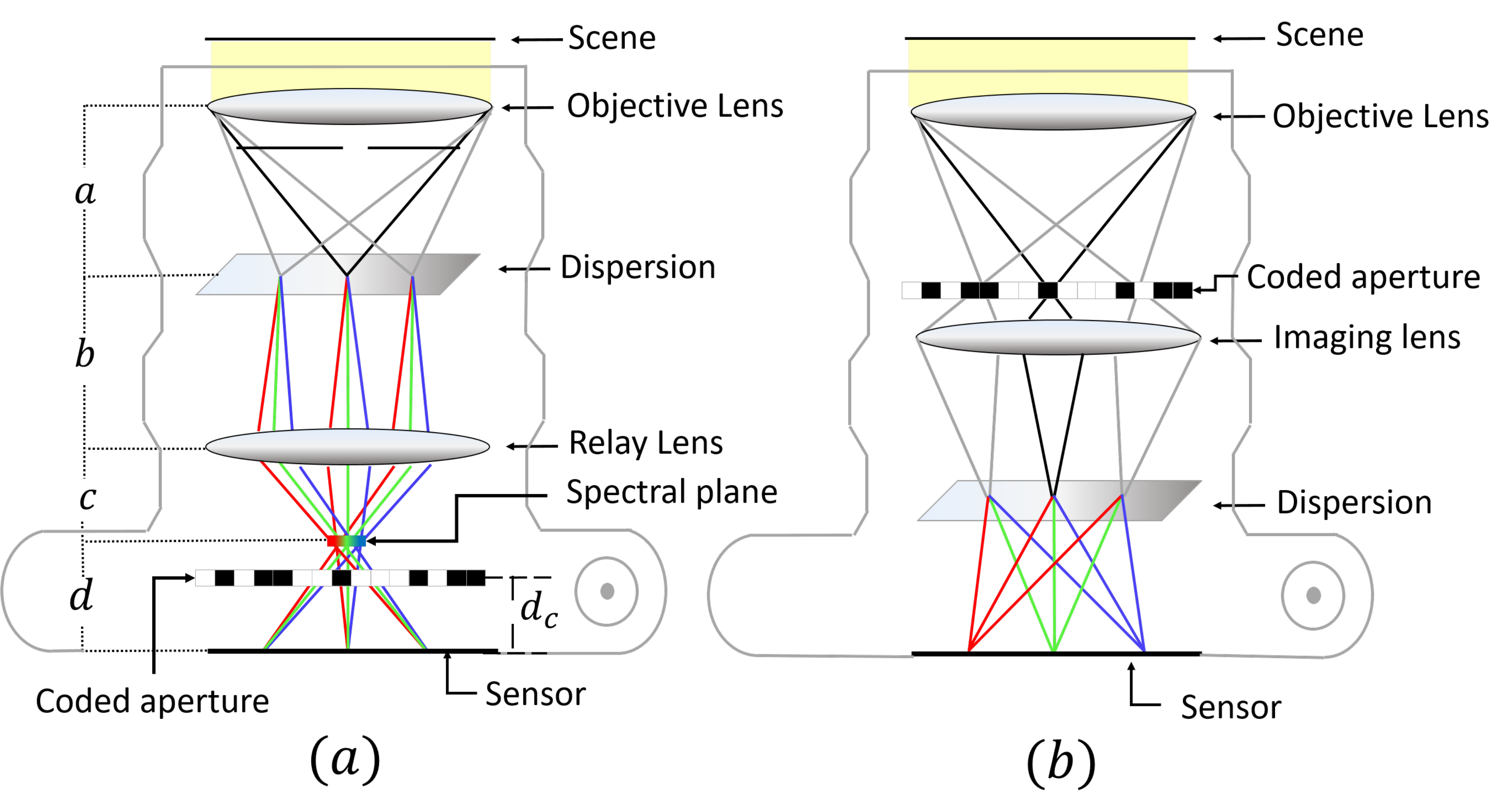}
 	\caption{Ray tracing diagram for (a) SSCSI, where the coded aperture is placed between the spectral plane and the sensor and (b) Single Disperser CASSI, where the coded aperture is located before the spectral dispersion.}
 	\label{set_up_2}
 \end{figure}

 Recently, Lin et al. \cite{lin} proposed a new optical architecture called \textit{Spatial Spectral Compressive Spectral Imager} (SSCSI). In this system, the spectral plane phenomena is exploited in order to carry out a three dimensional coding of the hyperspectral scene using a binary on-off mask. The spectral plane, see Fig. \ref{set_up_2}(a), is defined as the location where all the rays coming from the same wavelength meet each other in a single spatial point \cite{mohan}. The SSCSI was first introduced as an attempt to replicate the functionality of the dual disperser CASSI, avoiding the  calibration challenges, \cite{dual} and \cite{multicameras}. Coding in three dimensions, without the need of a colored mask, leads to high quality reconstructions while maintaining a low fabrication cost. Figure \ref{set_up_2} depicts the ray tracing diagram of the SSCSI and the single disperser CASSI. The SSCSI system can be seen in Fig. \ref{set_up_2}(a), where the spatial-spectral coding process takes place after the spectral dispersion, and an in-focus image impinges on the sensor. In CASSI (see Fig. \ref{set_up_2}(b)), the coded aperture modulates an in-focus image of the scene and the dispersion occurs after the code modulation.
 \\

 Some comparisons of the SSCSI with the different compressive spectral imagers have been done recently (Mar{\'\i}n  et al. in \cite{marin} and Medina-Rojas et al. in \cite{medina}).  Nevertheless, a rigorous analysis of the SSCSI has not been developed, particularly on the resolution limits and the various parameters affecting the quality of the recovered scene. This paper provides such analysis with respect to the coded aperture and detector pitch sizes.  
 Based on the continuous model studied in \cite{lin}, a discrete expression of the sensor measurements is obtained. It is shown in section V-B and V-C, that spatial super-resolution is achievable, and a spectral zooming process can be done by displacing the coded aperture towards the spectral plane. Here, the term spectral zooming refers to the increment of the number of resolvable spectral bands on the recovered scene. This model is compared, by numerical simulations, against CASSI and colored CASSI in terms of spatial and spectral accuracy in the reconstruction. Additionally, a set of testbed experiments is performed in order to
 validate the underlying concepts. Some preliminary results of this research are described in \cite{Salazar_1} and \cite{Salazar_2}.
 
This paper is organized as follows. In Section II, a description of the continuous model of SSCSI is presented. Section III presents the discretization of this model, offering a detailed description of the operators involved in the process. The structure of the sensing matrix is described in Section IV and the numerical simulations are presented in Section V. Sections VI and VII contain the experimental results and the conclusions, respectively.

\section{Sensing phenomena in SSCSI}
 
\begin{figure}[H]
	\centering
	\includegraphics[scale=0.15]{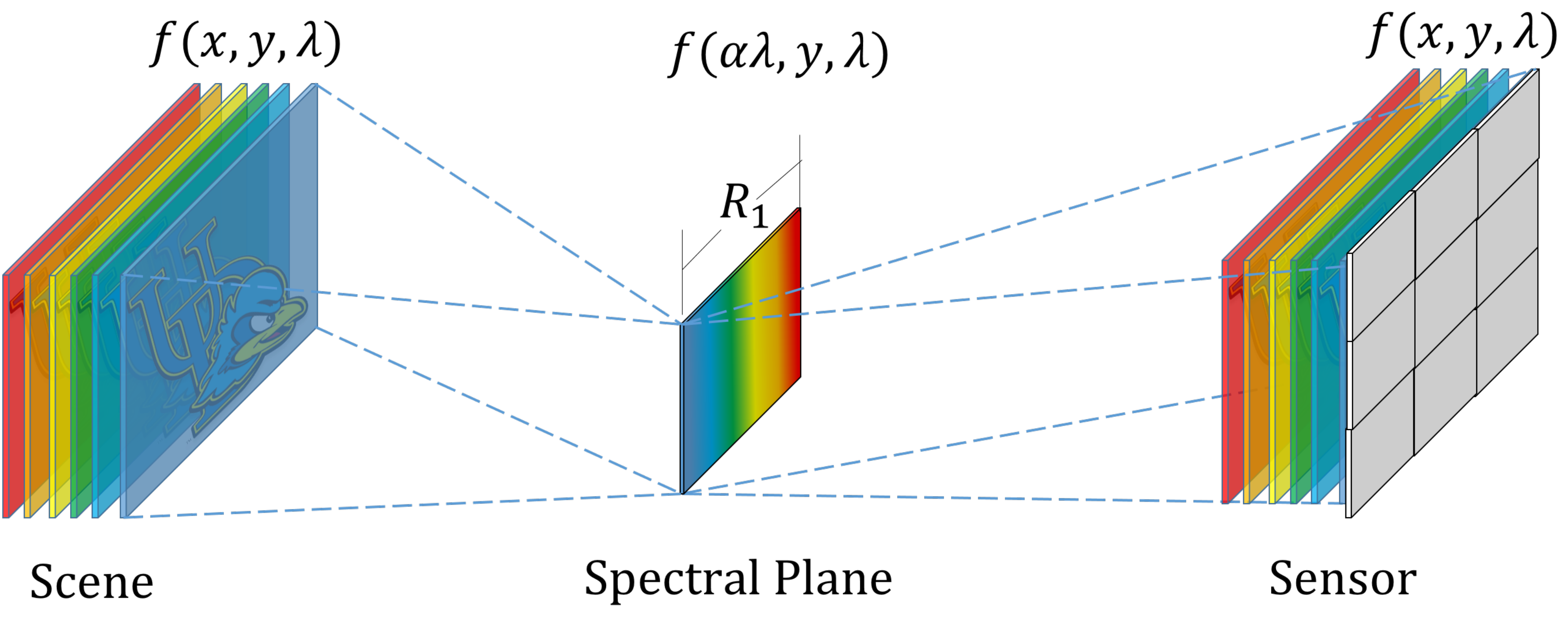}
	\caption{Ray propagation of the spectral scene onto the spectral plane and into the sensor. Here, $\textsl{g}(x,y)$ is the continuous model representation of the information captured at the sensor. The elements $(m=0,n=0)$ and $(m=2,n=0)$ of the detector array, are pointed out for clarification purposes.}
	\label{scheme}
\end{figure}

\begin{figure*}
	\centering
	\includegraphics[scale=0.65]{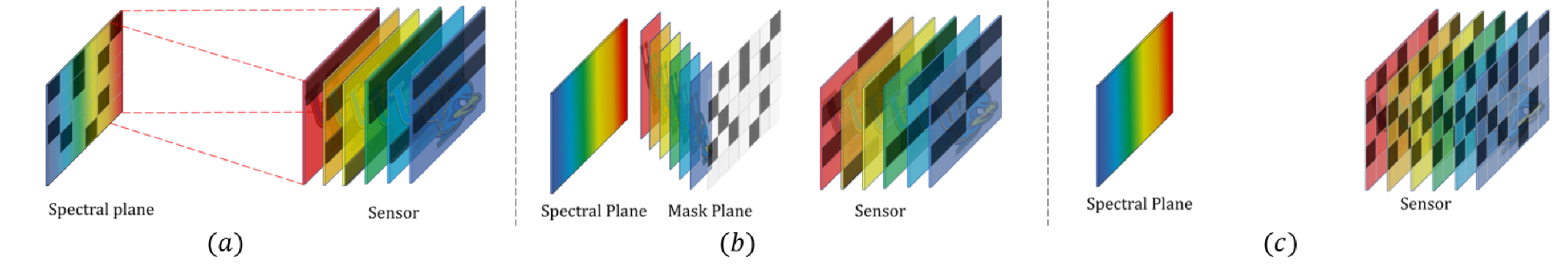}
	\caption{(a) Coding at the spectral plane ($s=1$), and the coded field at the detector. (b) Coding at $0 < s < 1$ and the coded field at the detector. (c) Coding when $s=0$ or the coded aperture is located at the detector.}
	\label{any_spec}
\end{figure*}

This section contains the description of the sensing process for the SSCSI, developed in \cite{lin}. The SSCSI optical architecture depicted in Fig. \ref{set_up_2}(a), makes use of  an objective lens to locate an image of the scene directly on the diffraction grating, where the spectral dispersion takes place. The rely lens is then used to get an in-focus image onto the sensor. Figure \ref{scheme} depicts the ray propagation of the scene onto the spectral plane and into the sensor. Notice how all rays from a same wavelength converge to a single line on the spectral plane. The width of that spectral plane, $R_1$, is given by the following expression \cite{mohan} 

\begin{equation}
R_{1}=\frac{bd}{b+d}\theta,
\label{spec}
\end{equation}
where $b$ is the distance between the grating and the relay lens and $d$ is the distance between the spectral plane and the sensor (see Fig. \ref{set_up_2}(a)). The variable $\theta$ is the grating dispersion angle. 
The parameter $s=d_c/d$ (see Fig. \ref{set_up_2}(a)), represents the position of the coded aperture with respect to the sensor. This normalized variable ($s=0$ when the coded aperture is on top of the sensor and $s=1$ when the coded aperture is on the spectral plane), strongly affects the coding process happening on the SSCSI. As depicted in Fig. \ref{any_spec}(a), when $s=1$ each resolvable spectral band will be coded by a different column of the mask and a stretched version of the coded aperture will impinge on the sensor. When $0<s<1$ (see Fig. \ref{any_spec}(b)) a similar process occurs, but in this case each resolvable band will be coded by a bigger portion of the coded aperture. When $s=0$, on the other hand (see Fig. \ref{any_spec}(c)), all the spectral information will be coded by the same coded aperture pattern. As it will be described shortly, the parameter $s$ plays a crucial role in the resolution limits of SSCSI. The sensing process of the hyperspectral scene after being coded at a given $s$, is characterized by the following expression \cite{lin}
 \begin{equation}
 \textsl{g}(x,y)=\int_{\Lambda}T(x(1-s)+s\alpha\lambda,y)f(x,y,\lambda)d\lambda,
 \label{gen}
 \end{equation}
where $\textsl{g}(x,y)$ is the continuous model representation of the information captured at the sensor (see Fig. \ref{scheme}), $T(x,y)$ is the coded aperture function, $f(x,y,\lambda)$ is the continuous representation of the hyperspectral scene, $\alpha$ is a parameter that indicates the position of a given wavelength on the spectral plane (see Appendix D for more details), and  $\Lambda$ is the spectral range of interest, typically going from $400$nm to $700$nm \cite{lin}. Notice how the term $s$ appears on the argument of $T(x,y)$, which reflects the dependency of the coding process on that parameter. In this expression, it is assumed that the aperture of the objective lens is an infinitesimally small pinhole; however, in practice, an aperture with finite dimensions will introduce blurring effects as it will be explained in Section VI.
Given that the sensor is composed of a discrete array of elements, a discretization process of Eq. (\ref{gen}) is a crucial step in understanding the physical phenomena in terms of the real captured information; this will be done in the following section.


\section{Discretization of the sensing phenomena}
This section contains the steps followed in order to find a discrete expression for the captured data in the SSCSI. The discretization is performed based on the analysis of the product of the rectangular functions that represent both the coded aperture and the detector. The spatial resolution limits are first analyzed by assuming a single wavelength scene. This leads to a discrete measurement expression that is posteriorly extended to a multiple wavelength hyperspectral scene, by defining the spectral resolution limits on the SSCSI. 

Consider the $(m,n)^{th}$ pixel detector, where $0\leq m,n \leq N_d-1$ are integer and unitless values, and $N_d\times N_d$ are the dimensions of the sensor array. The captured data on $(m,n)$  can be written as follows
\begin{eqnarray}
\mathbf{g}_{m,n}=\int\limits_{x}\int\limits_{y}\int_{\Lambda}T(x(1-s)+s\alpha\lambda,y)f(x,y,\lambda) \times \nonumber \\ \mathrm{rect}\left(\frac{x}{\Delta_d}-m,\frac{y}{\Delta_d}-n\right) d\lambda dydx,
\label{disc_1}
\end{eqnarray}
where $\Delta_d$ is the pitch size of a single detector element and $\mathrm{rect}(x)$ is a rectangular function defined as follows

\begin{equation}
\mathrm{rect}(x)=\left\lbrace 
\begin{array}{ccc}
1 &\text{If}& 0\leq x 
\leq 1\\
0  & & \text{otherwise,}\\
\end{array}
\right.
\label{rect}
\end{equation}
Likewise, the coded aperture can be written as
\begin{equation}
T(x,y)=\sum_{m'}\sum_{n'}\mathbf{t}_{m',n'}\mathrm{rect}\left(\frac{x}{\Delta_c}-m',\frac{y}{\Delta_c}-n'\right),
\label{disc_2}
\end{equation}
where $\mathbf{t}_{m',n'}\in\{ 0,1\}\quad\forall \ (m',n')$, $0\leq m',n' \leq N_c-1$  are integer and unitless values, and $N_c \times N_c$ are the dimensions of the coded aperture mask. The variable $\Delta_c$ is the pitch size of the coded aperture. In this paper, the coded aperture full width is assumed to be the same as the detector full width, or $N_c\Delta_c=N_d\Delta_d$. Moreover, the detector and the mask are assumed to be totally aligned with respect to each other. Replacing (\ref{disc_2}) into (\ref{disc_1}) leads to
\begin{multline}
\mathbf{g}_{m,n}=\int\limits_x\int\limits_y\int_{\Lambda}\sum_{m'}\sum_{n'}\mathbf{t}_{m',n'}\\
\times f(x,y,\lambda)\mathrm{rect}\left(\frac{x}{\Delta_d}-m,\frac{y}{\Delta_d}-n\right) \\
\times \mathrm{rect}\left(\frac{x(1-s)+s\alpha\lambda}{\Delta_c}-m',\frac{y}{\Delta_c}-n'\right)d\lambda dydx.
\label{disc_3} 
\end{multline}

A discretization measurement model of the last equation implies first a rigorous analysis of the spatial and spectral resolution limits related to the SSCSI, which will be done in the next subsections.


\subsection{Spatial Resolution Limits}
Consider a spectral scene $f(x,y,\lambda_o)\delta(\lambda-\lambda_o)$ having just one spectral component, where $\delta(\lambda-\lambda_o)$ is a Delta Dirac function located at $\lambda=\lambda_o$. Equation (\ref{disc_3}) can be rewritten as
\begin{multline}
\mathbf{g}_{m,n}=\int\limits_x\int\limits_y\int\limits_\Lambda\sum_{m'}\sum_{n'}\mathbf{t}_{m',n'}\\
 	\times f(x,y,\lambda_o)\mathrm{rect}\left(\frac{x}{\Delta_d}-m,\frac{y}{\Delta_d}-n\right)  \\
 	\times \mathrm{rect}\left(\frac{x(1-s)+s\alpha\lambda}{\Delta_c}-m',\frac{y}{\Delta_c}-n'\right)\delta(\lambda-\lambda_o)d\lambda dydx.
 	\label{disc_4_add} 
\end{multline}
By solving the integral over $d\lambda$, Eq. (\ref{disc_4_add}) can be rewritten as
\begin{multline}
\mathbf{g}_{m,n}=\int\limits_x\int\limits_y\sum_{m'}\sum_{n'}\mathbf{t}_{m',n'}\\
\times f(x,y,\lambda_o)\mathrm{rect}\left(\frac{x}{\Delta_d}-m,\frac{y}{\Delta_d}-n\right)  \\
\times \mathrm{rect}\left(\frac{x(1-s)+s\alpha\lambda_o}{\Delta_c}-m',\frac{y}{\Delta_c}-n'\right)dydx.
\label{disc_4} 
\end{multline}

The minimum spatially resolvable feature, given by Eq. (\ref{disc_4}) can be determined by establishing the intersecting region of the two rectangular functions in that expression. As it will be explained, this region is going to depend on the size of a coded aperture pixel  projected onto the sensor.
\subsubsection{When $\frac{\Delta_c}{1-s} \leq \Delta_d$}
\begin{figure}[H]
	\centering
	\includegraphics[scale=0.5]{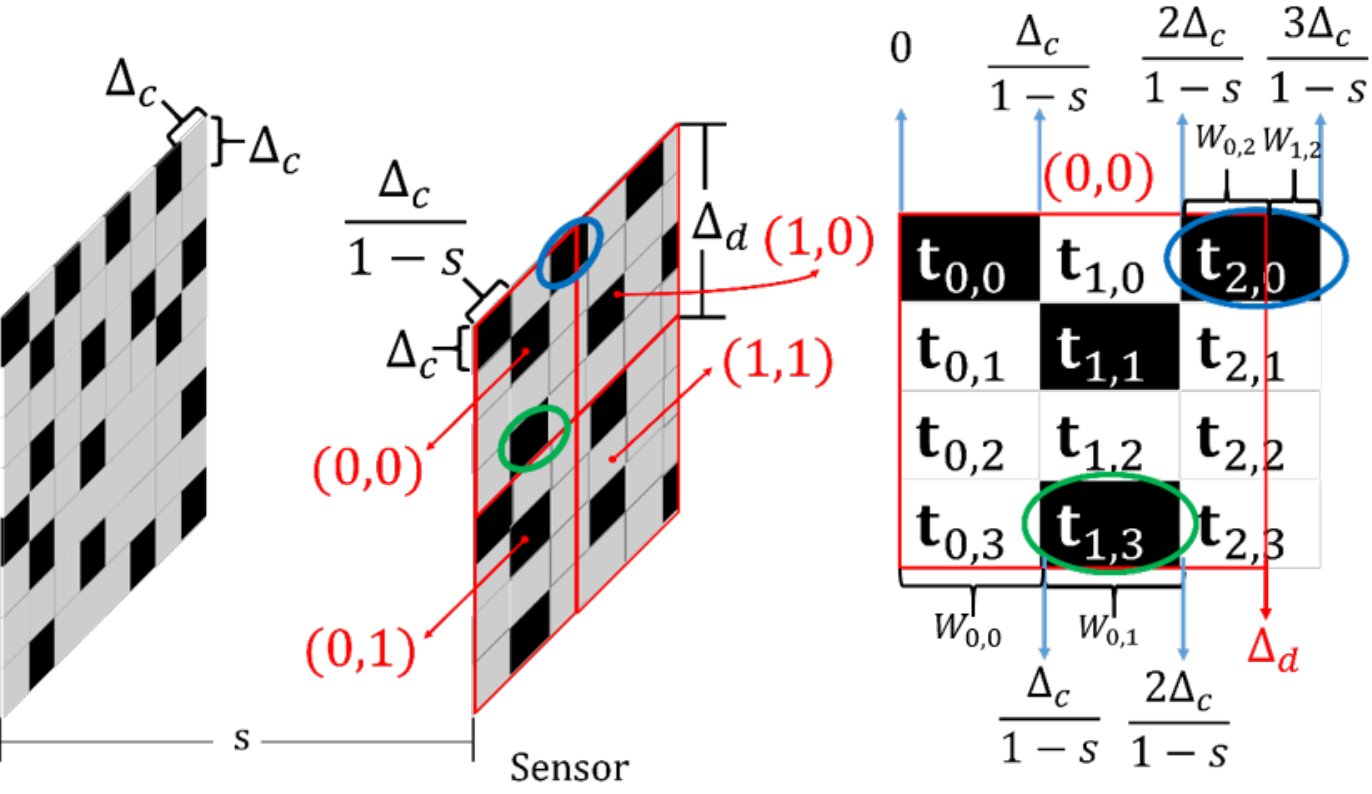}
	\caption{Left: Original Coded aperture at a given position $s$ and projection of the coded aperture onto the sensor. Here $\frac{\Delta_c}{1-s} \leq \Delta_d$. The red squares represent  the sensor elements $(0,0)$, $(0,1)$, $(1,0)$ and $(1,1)$. As depicted, a coded aperture pixel of size $\Delta_c \times \Delta_c$ is mapped into a $\frac{\Delta_c}{1-s} \times \Delta_c$ rectangle on the detector. Right: Front view of the sensor element (0,0) and illustration of the $W$ parameters of Eq. (\ref{disc_5}). The $W$ parameters represent the percentage of a coded aperture pixel impinging on a given detector element. Notice, for example, that one of the portions  of  $\mathbf{t}_{2,0}$ impinges on the $(1,0)^{th}$ sensor element; this portion is represented by $W_{1,2}$ (see Appendix A for full explanation).}
	\label{gen_mod}
\end{figure}

Figure \ref{gen_mod} left illustrates the coded aperture at a given position $s$ and its projection onto the sensor; the red lines delimit the detector elements. As depicted, a single coded aperture pixel of dimensions $\Delta_c \times \Delta_c$ is mapped into a $\frac{\Delta_c}{1-s} \times \Delta_c$ rectangle on the detector. This area defines the attainable spatial resolution of the recovered scene, as long as $\frac{\Delta_c}{1-s} \leq \Delta_d$. In this paper, it is assumed that $\Delta_c\leq\Delta_d$ and that the ratio $\Delta_d/\Delta_c$ is an integer greater than or equal to 1 (see Appendix E for $\Delta_c>\Delta_d$); this last condition implies that the mismatch between the coded aperture and the detector only applies to one axis (see Fig. \ref{gen_mod}, blue circled region). 

\begin{figure}[H]
	\centering
	\includegraphics[scale=0.4]{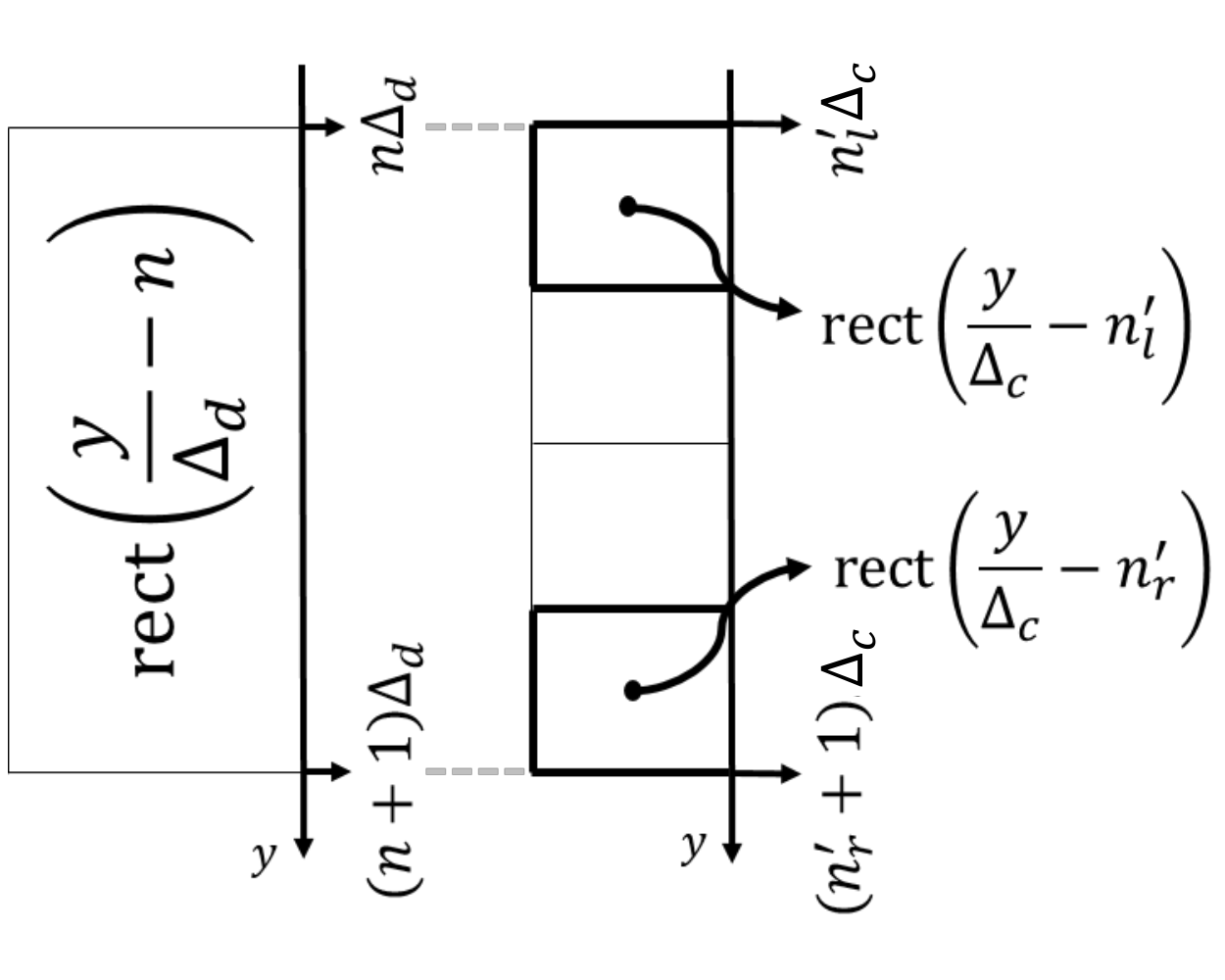}
	\caption{Overlapping area in the $y$ axis of the two rectangular functions given in Eq. (\ref{disc_4}), for a given $n$. Here, it is assumed that the ratio  $\Delta_d/\Delta_c$ is an integer greater than or equal to 1. Notice that $n'_l\Delta_c=n\Delta_d$ and $(n'_r+1)\Delta_c=(n+1)\Delta_d$.}
	\label{annex_eq_2}
\end{figure}

\begin{figure}[H]
	\centering
	\includegraphics[scale=0.3]{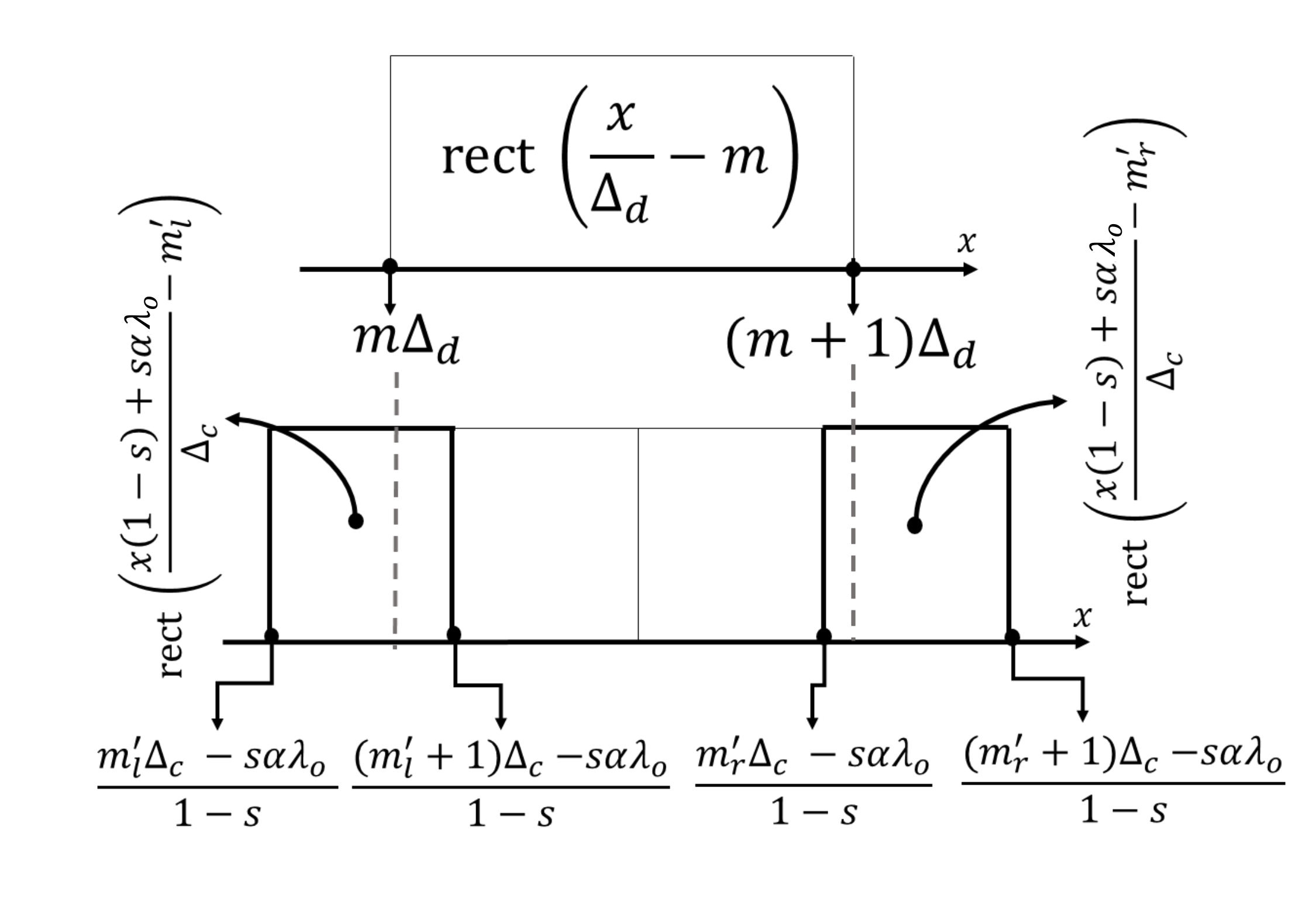}
	\caption{Overlapping area in the $x$ axis of the two rectangular functions in Eq. (\ref{disc_4}), for a given $m$. Notice that $\frac{m'_l\Delta_c-s\alpha\lambda_o}{1-s}\leq m\Delta_d\leq\frac{(m'_l+1)\Delta_c-s\alpha\lambda_o}{1-s}$ and $\frac{m'_r\Delta_c-s\alpha\lambda_o}{1-s}\leq (m+1)\Delta_d\leq\frac{(m'_r+1)\Delta_c-s\alpha\lambda_o}{1-s}$.}
	\label{annex_eq}
\end{figure}

Let $(m_l',n'_l)$ and $(m'_r,n'_r)$ be the upper-leftmost and lower-rightmost coded aperture elements impinging on the $(m,n)^{th}$ detector element. The overlapping of the two rectangular functions in Eq. (\ref{disc_4}) can be seen in Figs. \ref{annex_eq_2} and \ref{annex_eq}. From Fig. \ref{annex_eq_2}, it can be inferred that for the overlap to exist on the $y$ axis, the following condition must hold 
 \begin{equation}
n\frac{\Delta_d}{\Delta_c} \leq n' \leq (n+1)\frac{\Delta_d}{\Delta_c}-1,
\label{eq_n}
\end{equation}
where the limits in the last inequality come from solving $n'_l\Delta_c =n\Delta_d $ and $(n'_r+1)\Delta_c =(n+1)\Delta_d$ for $n'_l$ and $n'_r$ respectively. Likewise, it can be shown based on Fig. \ref{annex_eq}, that for the two rectangular functions in Eq. (\ref{disc_4}) to overlap in the $x$ axis, the next condition must hold 
 \begin{multline}
\left \lfloor \frac{(m)\Delta_d(1-s)+s\alpha\lambda_o}{\Delta_c} \right \rfloor \leq m' \leq \\
\left \lfloor \frac{(m+1)\Delta_d(1-s)+s\alpha\lambda_o}{\Delta_c} \right \rfloor,
\label{eq_m}
\end{multline}
 where $\lfloor \cdot \rfloor$ is the floor operator. The limits in the last inequality come from solving $\frac{m'_l\Delta_c-s\alpha\lambda_o}{1-s}\leq m\Delta_d\leq\frac{(m'_l+1)\Delta_c-s\alpha\lambda_o}{1-s}$
and $\frac{m'_r\Delta_c-s\alpha\lambda_o}{1-s}\leq (m+1)\Delta_d\leq\frac{(m'_r+1)\Delta_c-s\alpha\lambda_o}{1-s}$ for $m'_l$ and $m'_r$ respectively and considering that $m'_l$ and $m'_r$ are integer indexes. Given that the minimum resolvable spatial feature is of dimensions $\frac{\Delta_c}{(1-s)}\times \Delta_c$, and that the overlapping regions of the rectangular functions in Eq. (\ref{disc_4}) are defined in (\ref{eq_n}) and (\ref{eq_m}), the captured information in the $(m,n)^{th}$ detector element, $\mathbf{g}_{m,n}$, can be represented as


\vspace{-5mm}

\begin{equation}
\mathbf{g}_{m,n}=\sum_{m'=m'_l}^{m'_r} \ \ \sum_{n'=n\frac{\Delta_d}{\Delta_c}}^{(n+1)\frac{\Delta_d}{\Delta_c}-1}W_{m,m'}\times \mathbf{t}_{m',n'} \times \mathbf{f}_{m',n',\lambda_o},
\label{disc_5}
\end{equation}
where $m'_l$ and $m'_r$ are the limits of the expression in (\ref{eq_m}) and the variable $\mathbf{f}$ is the datacube representation of the hyperspectral scene, being $\mathbf{f}_{m',n',\lambda_o}$ the information of the datacube at position $(m',n')$  and spectral band $\lambda_o$; $\mathbf{f}_{m',n',\lambda_o}$ is obtained by applying the integral operators in Eq. (\ref{disc_4}). Notice that the spatial dimensions of $\mathbf{f}$, using an $N_d \times N_d$ sensor array, are defined as $\left \lceil N_d\frac{\Delta_d}{\Delta_c/(1-s)}\right \rceil\times N_d \frac{\Delta_d}{\Delta_c}$, where $\lceil \cdot \rceil$ is the ceiling operator. The parameter $W_{m,m'}$ is the fraction of the coded aperture element  $\mathbf{t}_{m',n'}$ impinging on the $(m,n)^{th}$ detector element, and it is defined as follows (see Appendix A for full derivation),
\begin{equation}
W_{m,m^{'}}=\left\lbrace 
\begin{array}{ccc}
\frac{\frac{(m'+1)\Delta_c-s\alpha\lambda_o}{1-s}-(m)\Delta_d}{\Delta_c/(1-s)} &\text{If}& m^{'}=m_{l}^{'}\\
\frac{(m+1)\Delta_d- \frac{m'\Delta_c-s\alpha\lambda_o}{1-s}}{\Delta_c/(1-s)}  &\text{If}& m^{'}=m_{r}^{'}\\
1   & \text{If}& m_{l}^{'}<m^{'}<m_{r}^{'}\\
0 & & \text{otherwise}.
\end{array}
\right.
\label{W_1}
\end{equation}
A similar analysis was previously done  for the CASSI system by Galvis et al. \cite{galvis}, in order to model the mismatching between the coded aperture and the sensor array. As an example, take the $(0,0)^{th}$ sensor element in Fig. \ref{gen_mod} right; the captured information in this element will be equal to $\mathbf{g}_{0,0}=\sum_{m'=0}^{2} \ \ \sum_{n'=0}^{3}W_{0,m'}\times \mathbf{t}_{m',n'} \times \mathbf{f}_{m',n',\lambda_o}$. In this particular case $W_{0,0}=1$ and $W_{0,1}=1$.
\subsubsection{When $\frac{\Delta_c}{1-s} > \Delta_d$}
\begin{figure}
	\centering
	\includegraphics[scale=0.09]{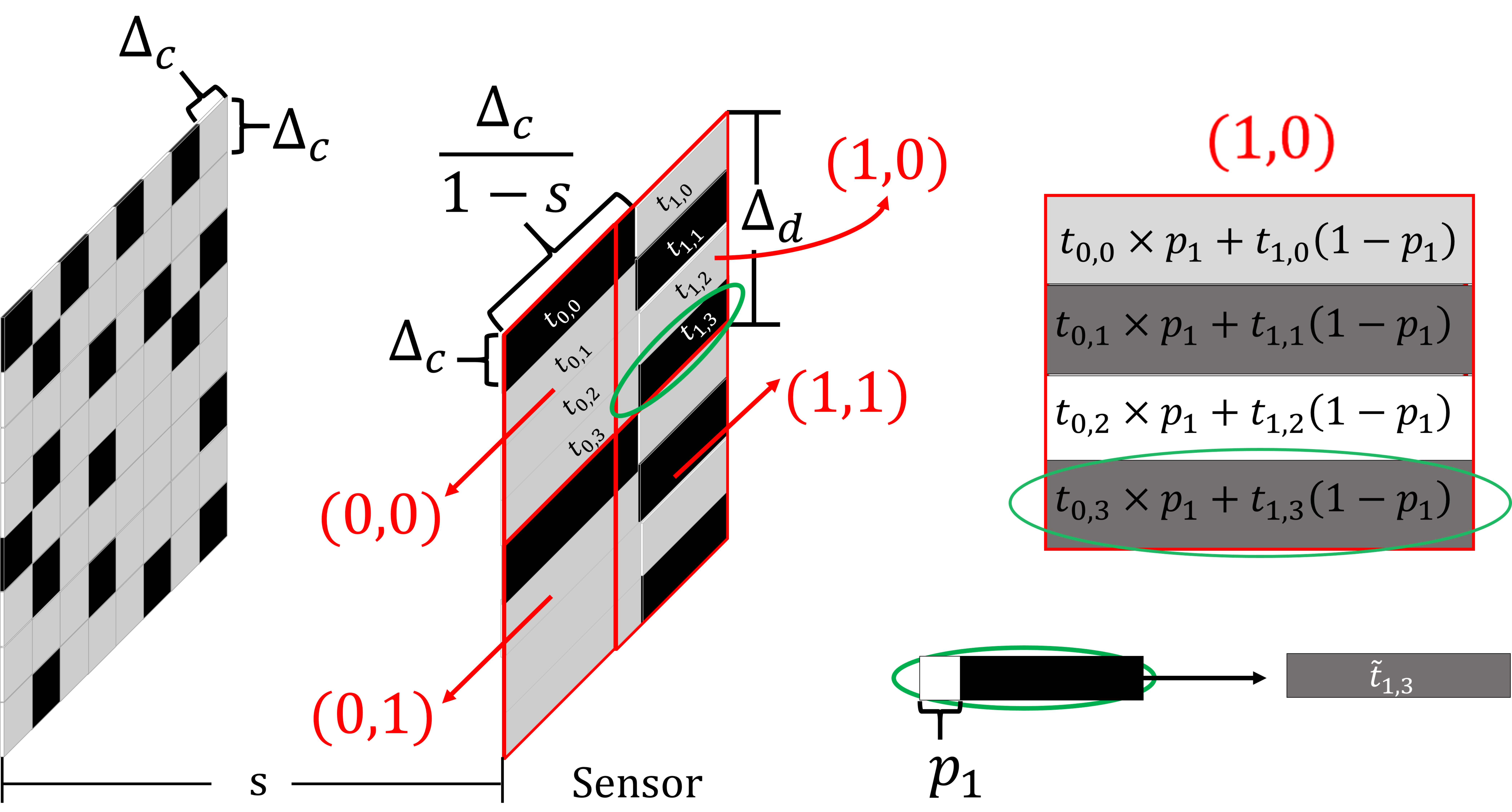}
	\caption{Left: Coded aperture at a given position $s$ and projection of the coded aperture onto the sensor. Here $\frac{\Delta_c}{1-s} > \Delta_d$. The red squares represent  the sensor elements $(0,0)$, $(0,1)$, $(1,0)$ and $(1,1)$. As depicted, a coded aperture pixel of size $\Delta_c \times \Delta_c$ is mapped into a $\frac{\Delta_c}{1-s} \times \Delta_c$ rectangle on the detector. Right: Effective coded aperture $\hat{\mathbf{t}}$ impinging in the $(1,0)^{th}$ sensor element. Notice, in the green circled region, that the variable $p_1$ indicates the area of the sensor occupied by $\mathbf{t}_{0,3}$.}
	\label{eq_exp_3}
\end{figure}

The projection of the coded aperture onto the sensor when $\frac{\Delta_c}{1-s} > \Delta_d$  is as shown in Fig. \ref{eq_exp_3} left. Unlike the last scenario, the attainable spatial resolution in the $x$ axis is limited not by $\frac{\Delta_c}{1-s}$ but by the detector pitch size $\Delta_d$; the minimum spatially resolvable feature is then of dimensions $\Delta_d \times \Delta_c$. As when $\frac{\Delta_c}{1-s} \leq \Delta_d$ , a coded aperture-sensor mismatch is present. To deal with this particularity, a synthetic effective pattern is generated by the combination of elements proportional to the areas, as illustrated by the green circle in Fig. \ref{eq_exp_3} right.\\
\indent Given the $(m',n')^{th}$ coded aperture pixel and the $(m,n)^{th}$ detector element, the two rectangular functions in Eq. (\ref{disc_4}) intersect each other if $n'$ falls on the interval specified in (\ref{eq_n}) and $m'=\left \lfloor \frac{(m)\Delta_d(1-s)+s\alpha\lambda_o}{\Delta_c} \right \rfloor +1$ (see Appendix A). Taking into account that the minimum spatially resolvable feature is of dimensions $\Delta_d\times\Delta_c$, the captured information in a particular sensor element, $\mathbf{g}_{m,n}$, can be written as follows
\begin{multline}
\mathbf{g}_{m,n}=\sum_{n'=n\frac{\Delta_d}{\Delta_c}}^{(n+1)\frac{\Delta_d}{\Delta_c}-1} \left(\mathbf{t}_{m'-1,n'}\times p_m + \mathbf{t}_{m',n'}\times (1-p_m) \right) \\\times \mathbf{f}_{m,n',\lambda_o},
\label{disc_9}
\end{multline}
where $\mathbf{f}_{m,n',\lambda_o}$ is obtained by applying the integral operators in Eq. (\ref{disc_4}). Notice that the spatial dimensions of $\mathbf{f}$, using an $N_d \times N_d$ sensor array, are defined as $N_d\times N_d \frac{\Delta_d}{\Delta_c}$. The expression $\tilde{\mathbf{t}}_{m,n'}=\mathbf{t}_{m'-1,n'}\times p_m + \mathbf{t}_{m',n'}\times (1-p_m)$ represents the effective coded aperture, where $p_m$ indicates the percentage occupied by $\mathbf{t}_{m'-1,n'}$ on the $(m,n)^{th}$ detector element and is defined as follows (see Appendix A),

\begin{equation}
p_m=\left\lbrace
\begin{array}{ccc}
1 & \text{If} & (m+1)\Delta_d\leq\frac{m'\Delta_c-s\alpha\lambda_o}{(1-s)}\\
\frac{\frac{m'\Delta_c-s\alpha\lambda_o}{(1-s)}-(m)\Delta_d}{\Delta_d} & \text{If} &(m+1)\Delta_d > \frac{m'\Delta_c-s\alpha\lambda_o}{(1-s)}
\end{array}
\right. 
\label{p_2}
\end{equation}

Unlike $W_{m,m'}$, that represents a percentage value with respect to a coded aperture pixel, $p_m$ is a percentage value with respect to the detector pitch size $\Delta_d$. As an example, take the $(1,0)^{th}$ sensor element in Fig. \ref{eq_exp_3} right. The captured information in this particular case will be equal to: $\mathbf{g}_{1,0}=\sum_{n'=0}^{3} \left(\mathbf{t}_{0,n'}\times p_1 + \mathbf{t}_{1,n'}\times (1-p_1) \right) \times \mathbf{f}_{1,n',\lambda_o}.$\\

Now that the spatial resolution limits were defined, an analysis to determine the minimum resolvable spectral band on the SSCSI must be developed in order to extend the discretization models, given in Eqs. (\ref{disc_5}) and (\ref{disc_9}), to the multi-spectral case.

\subsection{ Spectral Resolution Limits}
\begin{figure}
	\centering
	\includegraphics[scale=0.36]{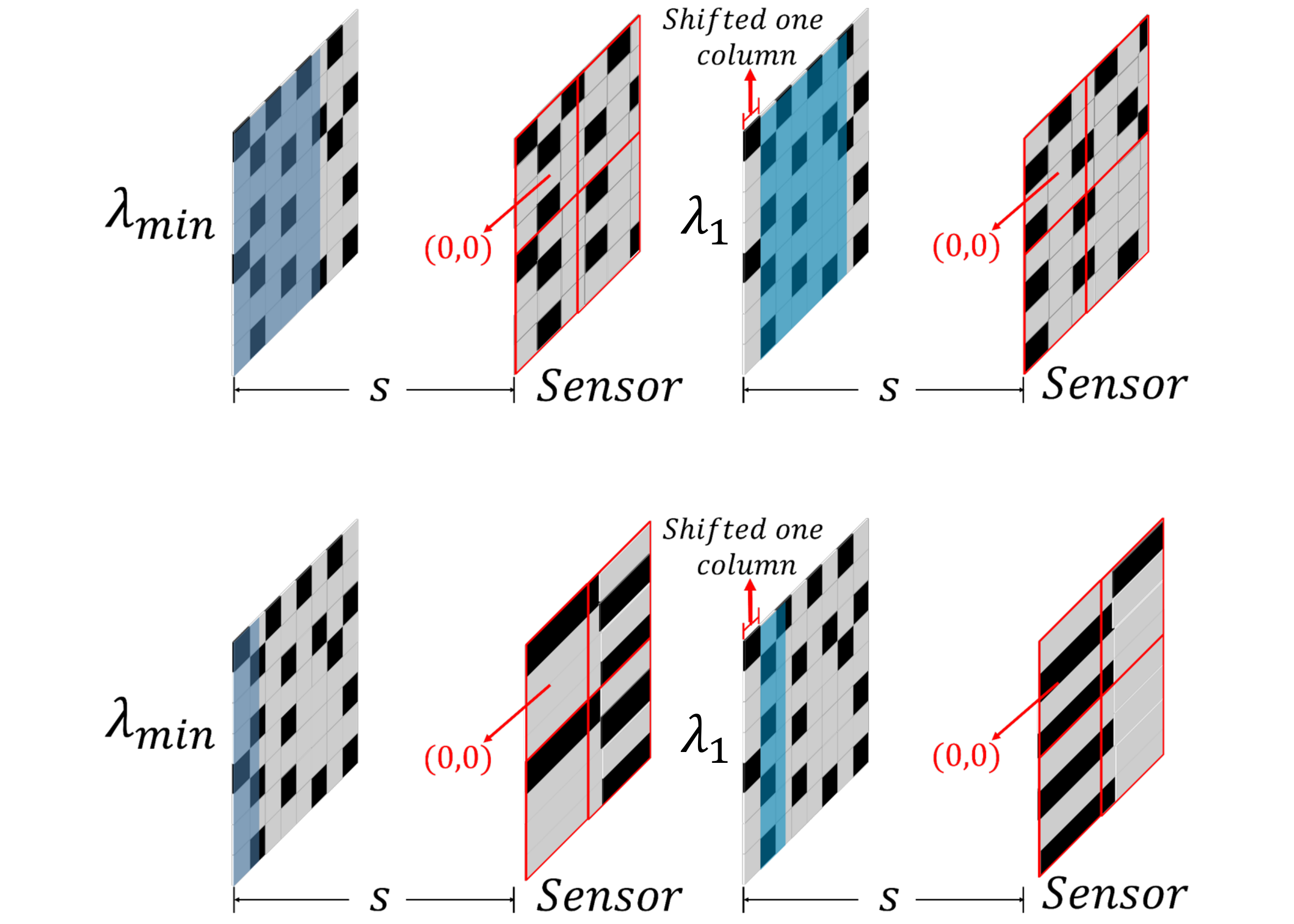}
	\caption{Top: Coding process, at a given $s$, for two adjacent and resolvable bands $\lambda_{min}$ and $\lambda_1=\lambda_{min}+\frac{\Delta_c}{s\alpha}$, when $\Delta_c/(1-s) \leq \Delta_d$. Bottom: Coding process, at a given $s$, for two adjacent and resolvable bands $\lambda_{min}$ and $\lambda_1=\lambda_{min}+\frac{\Delta_c}{s\alpha}$, when $\Delta_c/(1-s) > \Delta_d$. Notice that, in both cases $\lambda_{min}$ impinges on the leftmost side of the coded aperture. The dispersion between $\lambda_1$ and $\lambda_{min}$ is  equal to $s\alpha(\lambda_1-\lambda_{min})=\Delta_c$. This means they are coded by the same coded aperture but shifted by one column. }
	\label{spec_resol}
\end{figure}

A similar process to the one done to calculate the spatial resolution must be done to find an expression for the spectral resolution in SSCSI. In this case, a multispectral scene $f(x_o,y_o,\lambda)\delta(x-x_o,y-y_o)$ representing a single spatial location, where $\delta(x-x_o,y-y_o)$ is a two dimensional Delta Dirac function located at $(x=x_o, y=y_o)$,  is going to be considered. With this assumption, Equation (\ref{disc_3}) can be rewritten as follows
\begin{multline}
\mathbf{g}_{m,n}=\int\limits_x\int\limits_y\int_{\Lambda}\sum_{m'}\sum_{n'}\mathbf{t}_{m',n'} \\
\times f(x_o,y_o,\lambda)\mathrm{rect}\left(\frac{x}{\Delta_d}-m,\frac{y}{\Delta_d}-n\right) \\
\times \mathrm{rect}\left(\frac{x(1-s)+s\alpha\lambda}{\Delta_c}-m',\frac{y}{\Delta_c}-n'\right)\\
\times \delta(x-x_o,y-y_o)d\lambda dydx,
\label{spec_2_ADD} 
\end{multline}
By solving the integral over $dxdy$, Equation (\ref{spec_2_ADD}) can be rewritten as
\begin{multline}
\mathbf{g}_{m,n}=\int_{\Lambda}\sum_{m'}\sum_{n'}\mathbf{t}_{m',n'} \\
\times f(x_o,y_o,\lambda)\mathrm{rect}\left(\frac{x_o}{\Delta_d}-m,\frac{y_o}{\Delta_d}-n\right) \\
\times \mathrm{rect}\left(\frac{x_o(1-s)+s\alpha\lambda}{\Delta_c}-m',\frac{y_o}{\Delta_c}-n'\right)d\lambda,
\label{spec_2} 
\end{multline}
The attainable spectral resolution of the SSCSI is determined by the intersecting region of the two rectangular functions in Eq. (\ref{spec_2}). It can be shown that when  $\Delta_c/(1-s) \leq \Delta_d$, that region has a length of $\Delta_{\lambda}=\frac{\Delta_c}{s\alpha}$ for certain values of $s$ (see Appendix B). Therefore, the maximum number of resolvable spectral bands is given by
\begin{equation}
L=\left \lceil \frac{s\alpha(\lambda_{max}-\lambda_{min})}{\Delta_c} \right \rceil,
\label{bands}
\end{equation}
where $\lambda_{max}-\lambda_{min}=\vert \Lambda\vert$ is the spectral range of interest. Notice that the term $s\alpha(\lambda_{max}-\lambda_{min})$ indicates the spectral dispersion of the datacube for a given position $s$. Let  $\lambda_{k+1}=\lambda_{min}+\frac{\Delta_c}{s\alpha}(k+1)$ and $\lambda_{k}=\lambda_{min}+\frac{\Delta_c}{s\alpha}(k)$ be two wavelengths, such that $\lambda_{k+1}-\lambda_{k}=\frac{\Delta_c}{s\alpha}$. The dispersion between these two bands is given by $s\alpha(\lambda_{k+1}-\lambda_{k})=\Delta_c$. This means that two adjacent and resolvable bands will be coded by the same coded aperture but shifted by one column $(\Delta_c)$ as depicted in Fig. \ref{spec_resol}, top (for $k=0$).
Eq. (\ref{disc_5}) can therefore be  extended to the multispectral case as follows
\begin{equation}
\mathbf{g}_{m,n}=  \sum_{m'=m'_l}^{m'_r} \sum_{n'=n\frac{\Delta_d}{\Delta_c}}^{(n+1)\frac{\Delta_d}{\Delta_c}-1} \sum_{k=0}^{L-1}W_{m,m'} \mathbf{t}_{m'+k,n'} \mathbf{f}_{m',n',k},
\label{tot_7}
\end{equation}
where  $m'_l=\left \lfloor \frac{(m)\Delta_d(1-s)+s\alpha\lambda_{min}}{\Delta_c} \right \rfloor$, $m'_r=\left \lfloor \frac{(m+1)\Delta_d(1-s)+s\alpha\lambda_{min}}{\Delta_c} \right \rfloor$ and $W_{m,m'}$ is defined in Eq. (\ref{W_1}) (for $\lambda_o=\lambda_{min}$). The $k^{th}$ spectral band is defined here as the interval $[\lambda_{min}+\frac{\Delta_c}{s\alpha}k,\lambda_{min}+\frac{\Delta_c}{s\alpha}(k+1)]$; the shifting by one column between adjacent bands can be seen in the $``+k"$ term on the coded aperture array $\mathbf{t}$. The dimensions of the recovered datacube are given by $\left \lceil N_d\frac{\Delta_d}{\Delta_c/(1-s)} \right \rceil  \times N_d\frac{\Delta_d}{\Delta_c} \times L$, where $N_d \times N_d$ are the dimensions of the sensor array and $L$ is given in Eq. (\ref{bands}). As an example, take the sensor element $(0,0)$ in Fig. \ref{spec_resol}, top; the captured information in that element related to the spectral band $[\lambda_{min},\lambda_{min}+\frac{\Delta_c}{s\alpha}]$  will be given by $\mathbf{g}_{0,0}=  \sum_{m'=0}^{2} \sum_{n'=0}^{3} W_{0,m'} \mathbf{t}_{m',n'} \mathbf{f}_{m',n',0}$. Likewise, the captured information for the spectral band $[\lambda_{min}+\frac{\Delta_c}{s\alpha},\lambda_{min}+\frac{2\Delta_c}{s\alpha}]$  will be given by $ \mathbf{g}_{0,0}=  \sum_{m'=0}^{2} \sum_{n'=0}^{3} W_{0,m'} \mathbf{t}_{m'+1,n'} \mathbf{f}_{m',n',1}$. Notice that in this particular case, $W_{0,0}=1$ $W_{0,1}=1$. Since the actual shearing of the coded aperture seen at the sensor is approximated, this model can be called First order approximation. An analysis of the actual sheared coded aperture pattern  and the parallel with the proposed model can be found in Appendix C. 

When $\Delta_c/(1-s) >\Delta_d$, a similar analysis can be done to find that the attainable spectral resolution can be assumed as $\frac{\Delta_c}{s\alpha}$ (see Appendix B). Again, as when $\Delta_c/(1-s) \leq \Delta_d$, two adjacent and resolvable bands will be coded by the same coded aperture but shifted by one column, or $s\alpha(\lambda_{k+1}-\lambda_{k})=\Delta_c$, as depicted in Fig. \ref{spec_resol}, bottom (for $k=0$).~Eq. (\ref{disc_9}) can therefore, be extended to the multispectral case as follows
\begin{multline}
\mathbf{g}_{m,n}=\sum_{n'=n\frac{\Delta_d}{\Delta_c}}^{(n+1)\frac{\Delta_d}{\Delta_c}-1}\sum_{k=0}^{L-1}\left(\mathbf{t}_{m'+k-1,n'}\times p_m +\right. \\ 
\left. \mathbf{t}_{m'+k,n'}\times(1-p_m)\right) \mathbf{f}_{m,n',k},
\label{disc_10} 
\end{multline}
where the $k^{th}$ spectral band is defined as the interval $[\lambda_{min}+\frac{\Delta_c}{s\alpha}k,\lambda_{min}+\frac{\Delta_c}{s\alpha}(k+1)]$, $m'=\left \lfloor \frac{(m)\Delta_d(1-s)+s\alpha\lambda_{min}}{\Delta_c} \right \rfloor+1$ and $p_m$ is as defined in Eq. (\ref{p_2}) (for $\lambda_o=\lambda_{min}$). The shifting by one column between adjacent bands can be seen in the  $``+k"$ term on the coded aperture array $\mathbf{t}$. The dimensions of the recovered datacube are given by  $N_d \times N_d\frac{\Delta_d}{\Delta_c} \times L$, where $L$ is given in Eq. (\ref{bands}).
As an example, take the sensor element $(1,0)$ in Fig. \ref{spec_resol}, bottom. The captured information in that element related to the spectral band $[\lambda_{min},\lambda_{min}+\frac{\Delta_c}{s\alpha}]$ will be given by $\mathbf{g}_{1,0}=\sum_{n'=0}^{3}\left(\mathbf{t}_{0,n'}\times p_1 +\mathbf{t}_{1,n'}\times (1-p_1)\right)  \mathbf{f}_{1,n',0}$. Likewise,  the captured information for the spectral band $[\lambda_{min}+\frac{\Delta_c}{s\alpha},\lambda_{min}+\frac{2\Delta_c}{s\alpha}]$ will be given by $\mathbf{g}_{1,0}=\sum_{n'=0}^{3}\left(\mathbf{t}_{1,n'}\times p_1 +\mathbf{t}_{2,n'}\times (1-p_1)\right) \mathbf{f}_{1,n',1}$. Again, the present model is an approximation and an analysis of the actual sheared coded aperture can be seen in Appendix C.\\
Figure \ref{equi_1} shows the coded aperture projected onto the sensor and the equivalent coded aperture for two adjacent bands when $\frac{\Delta_c}{1-s} > \Delta_d$ and $\frac{\Delta_c}{1-s} \leq \Delta_d$. Notice that the recovered datacube might not be square, as the resolution in each spatial axis can be different. 

 \begin{figure}[h!]
 	\centering
 	\includegraphics[scale=0.43]{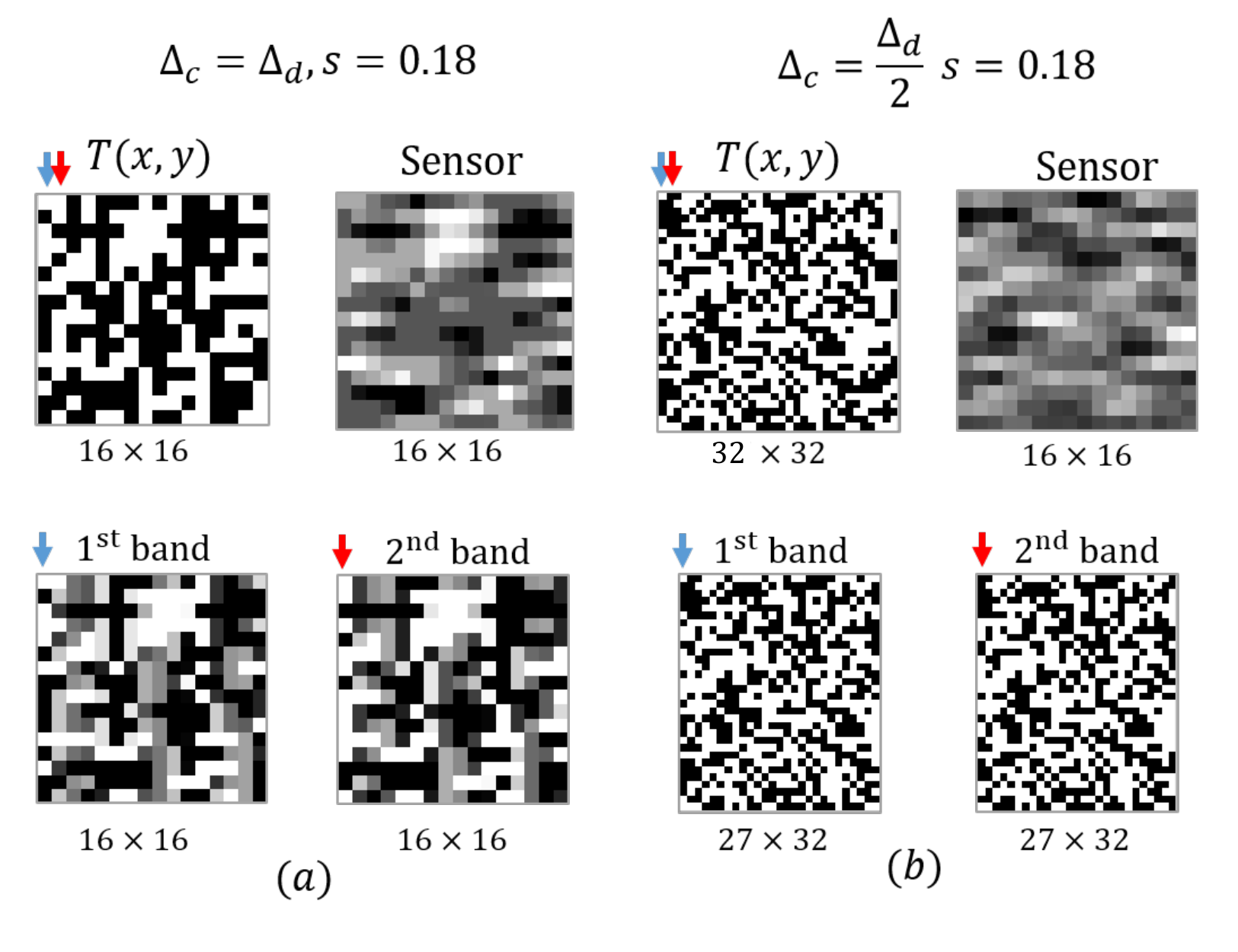}
 	\caption{(a) Top left: original coded aperture. Top right: coded aperture at the sensor. Bottom left and right: equivalent coded apertures that modulate the first and second band respectively. In this figure $\Delta_c/(1-s)>\Delta_d$. (b) Top Left: original coded aperture. Top right: coded aperture at the sensor. Bottom left and right: equivalent coded apertures that modulate the first and second band respectively. In this figure $\Delta_c/(1-s)<\Delta_d$. Notice how the  
 		pattern that codes the first band  starts at the first column of the original coded aperture $T(x,y)$ (blue arrow), while the patter that codes the second band starts at the second column of the original coded aperture $T(x,y)$ (red arrow).}
 	\label{equi_1}
 \end{figure}

The dependency of the number of resolvable bands on $\Delta_c$ and $s$, as stated in Eq. (\ref{bands}), allows to enhance the spectral resolution of the system by tunning these parameters. Increasing $s$, which means moving the coded aperture towards the spectral plane, is equivalent to do a zooming process over the spectral dimension of the datacube. However,  in experiments, placing the coded aperture at $s\neq0$ decreases the spatial resolution of the recovered scene. This will be further discussed in the experimental part of the present paper.\\

An alternative expression for the number of the resolvable bands can be seen in the following equation (see Appendix D for full derivation),

\begin{equation}
L= \left \lceil s\frac{N_d}{\Delta_c/\Delta_d}\beta \right\rceil,
\label{s_aprox}
\end{equation}

where $\beta=\alpha(\lambda_{max}-\lambda_{min})/(\Delta_cN_c)$ is the ratio between the width of the spectral plane and the coded aperture full width. If $\beta=1$, the spectral plane fully occupies the coded aperture width. The variable $\beta$ is related to $R_1$, given by Eq. (\ref{spec}); therefore, a convenient way to increase $\beta$ is choosing a diffraction grating element with high diffraction angle $\theta$.\\
\vspace{-5mm}


\section{Matrix Forward Model}
\begin{figure}
	\centering
	\includegraphics[scale=0.15]{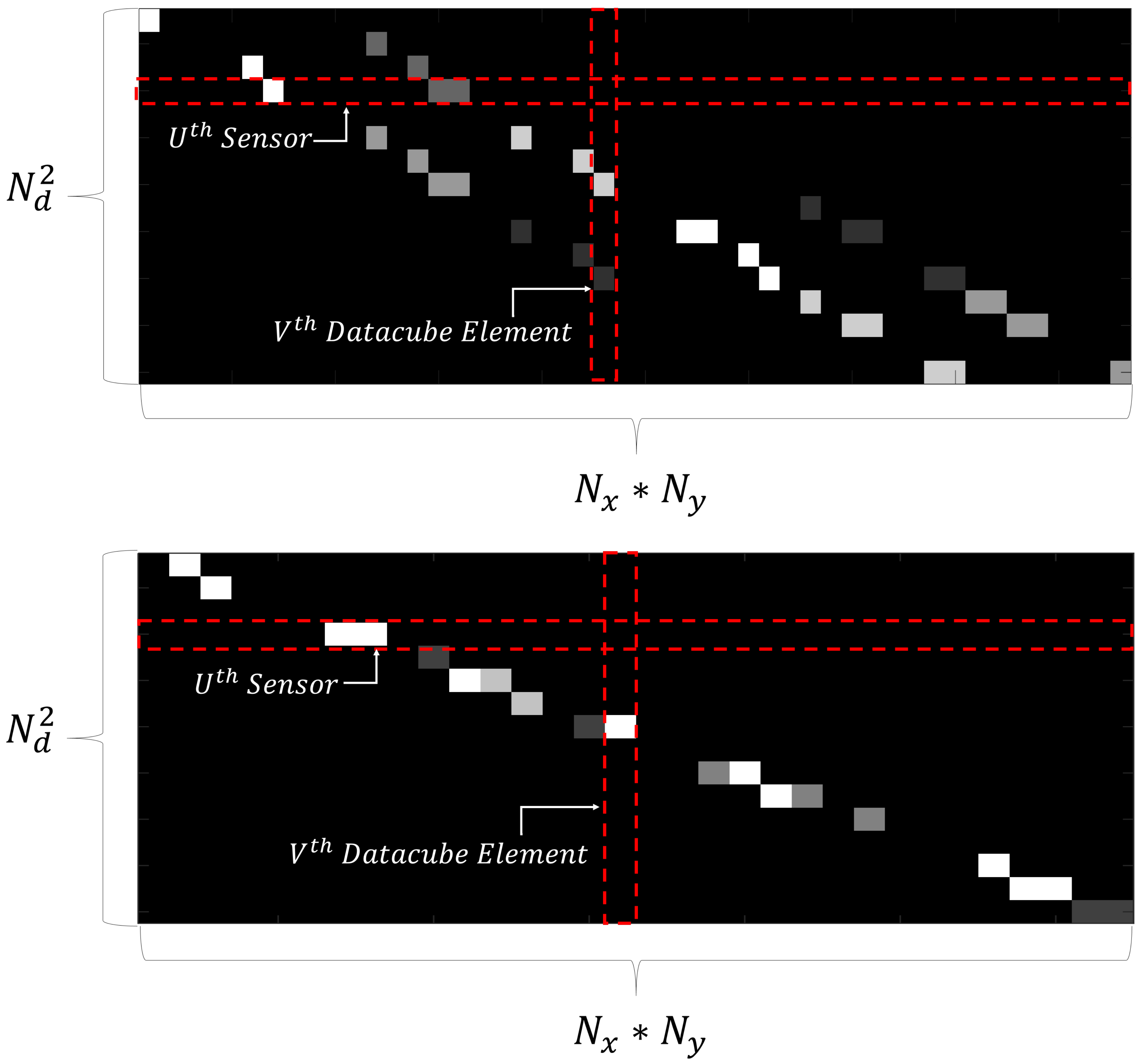}
	\caption{Sensing matrix $\mathbf{H}^{(q)}$, for a single shot and a $N_x \times N_y$ single wavelength datacube. Top: when $\frac{\Delta_c}{1-s} \leq \Delta_d$. Bottom: when $\frac{\Delta_c}{1-s} > \Delta_d$. Each row represents a sensor element and each column represents a datacube element.}
	\label{matrix_single2}
\end{figure}

 Given the discrete measurement model explained in the previous section, the sensing process can be expressed in a matrix form as follows
\begin{equation}
\vec{\mathbf{g}}^{(q)}=\mathbf{H}^{(q)} \boldsymbol{\Psi} \boldsymbol{\pi}\quad   q=1,2,...,Q,
\label{mfm}
\end{equation}
where $Q$ is the number of captured shots and $\vec{\mathbf{g}}^{(q)}$ is the column-wise version of the measurements of length $N_d^2$. The variable $\boldsymbol{\pi}$ is a column-wise sparse 
representation of the datacube $\mathbf{f}$ in a basis $\boldsymbol{\Psi}$. If the datacube to be recovered has dimensions $N_x \times N_y \times L$, then the length of $\boldsymbol{\pi}$ is $N_{x}N_{y}L$. The structured matrix $\mathbf{H}^{(q)}$ performs 
the coding 
process of the datacube; its dimensions are $N_d^2 \times N_{x}N_{y}L$.  Figure \ref{matrix_single2} shows the structure of $\mathbf{H}^{(q)}$  when $\frac{\Delta_c}{1-s} \leq \Delta_d$ 
and $\frac{\Delta_c}{1-s} > \Delta_d$ for a single wavelength datacube. Each row of $\mathbf{H}^{(q)}$ corresponds to a sensor element, while each column corresponds to a datacube component.
The values of the nonzero entries of the matrices are linked to Eqs. (\ref{tot_7}) and (\ref{disc_10}). Notice that one single sensor can measure several datacube elements.

\begin{figure}
	\centering
	\includegraphics[scale=0.28]{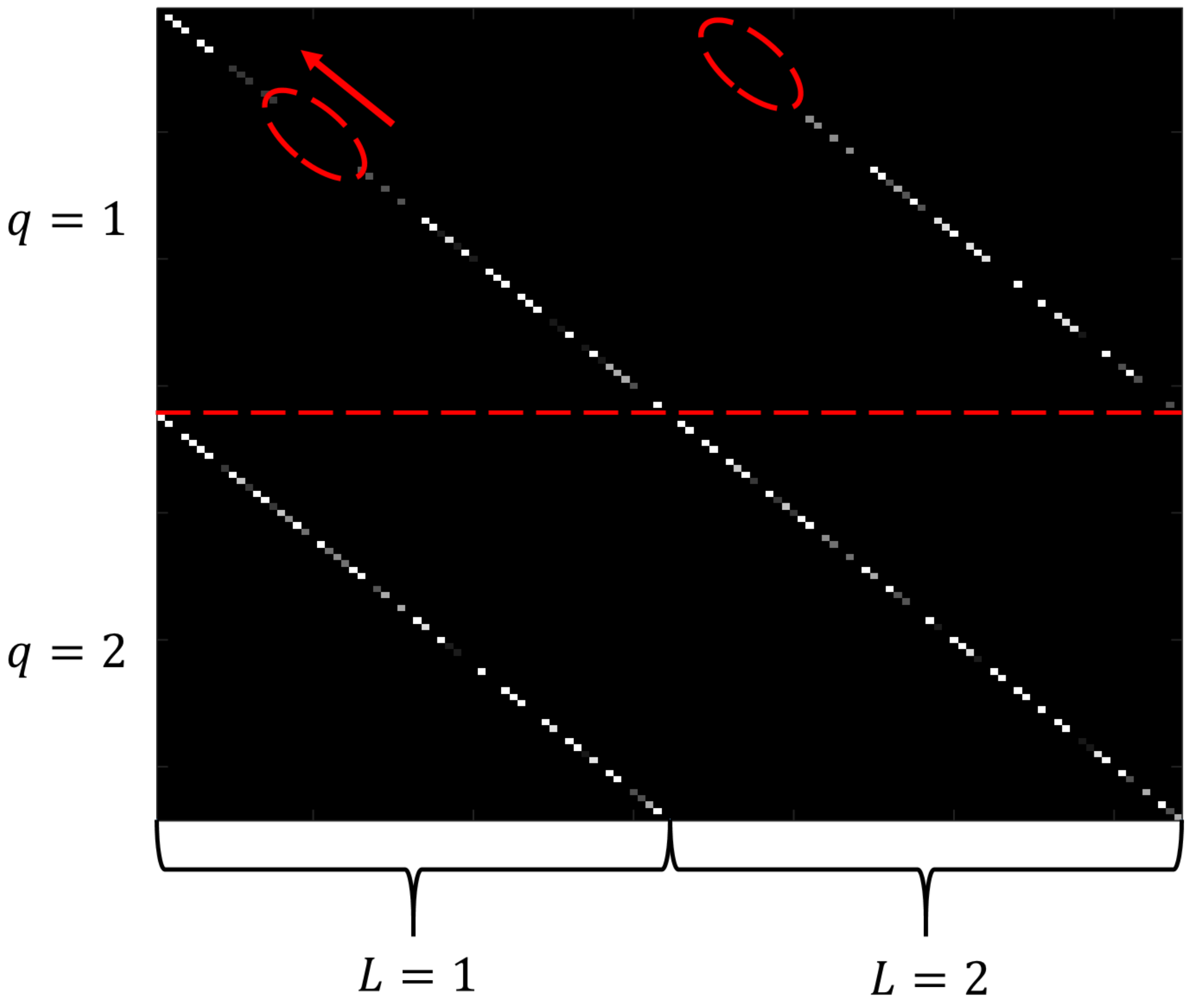}
	\caption{Sensing matrix for two shots and a two wavelengths datacube, where $\Delta_c=\Delta_d$, $s=0.18$ and $\beta=1$. The dashed red horizontal line separates the two different shots, $q=1$ and $q=2$.  The size of the implemented sensor and coded aperture is $8 \times 8$. The dashed red circles and red arrow indicate the dispersion process between bands.}
	\label{two_wav}
\end{figure}

For a multishot approach, the sensing process can be written as 
\begin{equation}
\vec{\mathbf{g}}= \mathbf{H}\boldsymbol{\Psi}\boldsymbol{\pi},
\end{equation}
 where $\mathbf{H}=\left[{\mathbf{H}^{(1)}}^{T},{\mathbf{H}^{(2)}}^{T}, \dots, {\mathbf{H}^{(Q)}}^{T}\right]^T$ is a structured matrix with dimensions $QN_d^2 \times N_{x}N_{y}L$, and $\vec{\mathbf{g}}=\left[ \vec{\mathbf{g}}^{(1)T},\vec{\mathbf{g}}^{(2)T}, \dots, \vec{\mathbf{g}}^{(Q)T}\right] ^T$ is a vector of length $QN_d^2$. The structure of $\mathbf{H}$ for $Q=2$ and $L=2$  can be seen in Fig. \ref{two_wav}. 
Unlike CASSI, the dispersion in the SSCSI is seen as the movement of the elements on the diagonals.  An example on how to ensemble $\mathbf{H}$ when $\Delta_d=\Delta_c$ and the coded aperture has the same dimensions than the detector $(N_d=N_c)$, can be seen in Algorithm 1. In this particular case, according to Eq. (\ref{eq_n}), $n'=n$. The computational complexity of this algorithm is  $\mathcal{O}(N^2LQ)$. When $\Delta_c/(1-s)<\Delta_d$, a similar process to the one specified in Algorithm 1 must be done. In this case, for every $(m,n)^{th}$ detector element, $n'$, $m'$ and $W_{m',m}$ is calculated using Eqs. (\ref{eq_n}) and (\ref{eq_m}) and (\ref{W_1}).


%
%
%
%

\begin{algorithm}
	\SetKwInOut{Input}{Input}
	\SetKwInOut{Output}{Output}
	
	\Input{$\mathbf{t}$,$\Delta$,$s$,$\alpha$,$\lambda_{min}$,$\lambda_{max}$,$N$}
	\Output{$\mathbf{H}$}
	\noindent Define $L$ as in Eq. (\ref{bands})\\
	\noindent $\mathbf{H}=[ \ ]$\\
	\For{$q=0:Q-1$}{
		\For{$m,n=0:N-1$}{
			
			\noindent Given $\lambda_o=\lambda_{min}$ and $m$, find $m'$ and $p_m$. \\
			\noindent $n'=n$. \\
			\For{$k=0:L-1$}
			{

				\noindent  $u=N\times m+n+1$.\\
				\noindent $v=N\times m+n+1+k\times N^2$.\\
				\noindent $\mathbf{H}^{(q)}(u,v)=\mathbf{t}^{q}_{m'+k-1,n}\times p_m+\mathbf{t}^q_{m'+k,n}\times(1-p_m)$\\         
			}
		}
		
		$\mathbf{H}=[\mathbf{H};\mathbf{H}^{(q)}]$
	}
	\caption{Computation of matrix $\mathbf{H}$ when $\Delta_d=\Delta_c=\Delta$ and $N_d=N_c=N$} 
\end{algorithm}

	
\section{Simulations}
This section contains the simulation results in order to test the SSCSI first order approximation. The model is first compared to CASSI in terms of spatial and spectral quality in the reconstructions. After that, the spatial super-resolution and the spectral zooming concepts are tested in simulations. The performance of the SSCSI at different values of $s$ is then analyzed. A last subsection regarding the coherence of the sensing matrix and the impact on the quality of the reconstruction, is included. A hyperspectral scene was captured at the Computational Imaging and Spectroscopy Laboratory at University of Delaware,  using  a visible monochromator between 451nm and 642nm and a CCD camera, to posteriorly implement it as a ground-truth in simulations. The sparsity basis was chosen as $\boldsymbol{\Psi}=\boldsymbol{\Psi}_{DCT} \otimes \boldsymbol{\Psi}_{W}$, where $\boldsymbol{\Psi}_{W}$ is the 2D Wavelet Symlet 8 basis, $\boldsymbol{\Psi}_{DCT}$ is the Discrete Cosine basis and $\otimes$ is the Kronecker product \cite{duartekronecker}. The reconstruction is obtained as the solution of
 \begin{equation}
 \underset{\boldsymbol{\pi}}{\mathrm{argmin}}\left(\left\Vert \vec{\textbf{g}}-\textbf{A}\boldsymbol{\pi} \right\Vert_2 + \tau\left\Vert\boldsymbol{\pi}\right\Vert_1 \right),
 \label{cs_eq}
 \end{equation}
 where $\left\Vert \cdot \right\Vert_2 $ is the $\ell_{2}$ norm,  $\left\Vert \cdot \right\Vert_1 $ is the $\ell_1$ norm, $\tau$ is a regularization parameter that controls the sparsity of the solution and $\textbf{A}=\mathbf{H}\boldsymbol{\Psi}$. 
The Gradient Projection for Sparse Reconstruction algorithm (GPSR) is used in order to solve Eq. (\ref{cs_eq}), as this algorithm has been shown to have optimal numerical performance and speed \cite{GPSR}.\\

\subsection{Comparison to CASSI and Colored CASSI}
\begin{figure}[h!]
	\centering
	\includegraphics[scale=0.4]{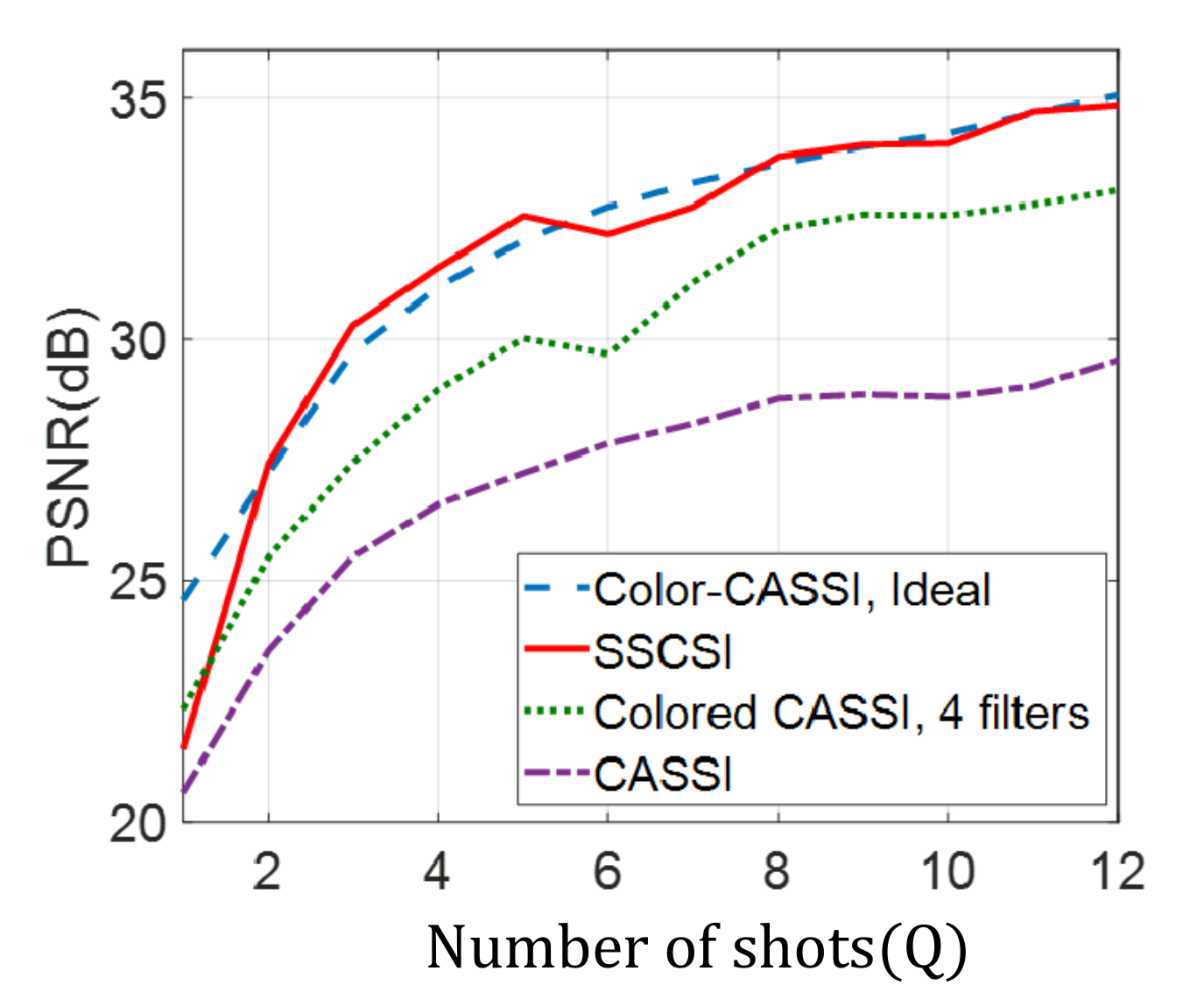}
	\caption{PSNR of the reconstruction as a function of the number of shots for the SSCSI, $s\approx0.1$,  CASSI, colored CASSI with 4 filters and Ideal Colored CASSI. All the reconstructions were done assuming $\Delta_c=\Delta_d$.}
	\label{cassi_v_sscsi}
\end{figure}

\begin{figure}[h!]
	\centering
	\includegraphics[scale=0.4]{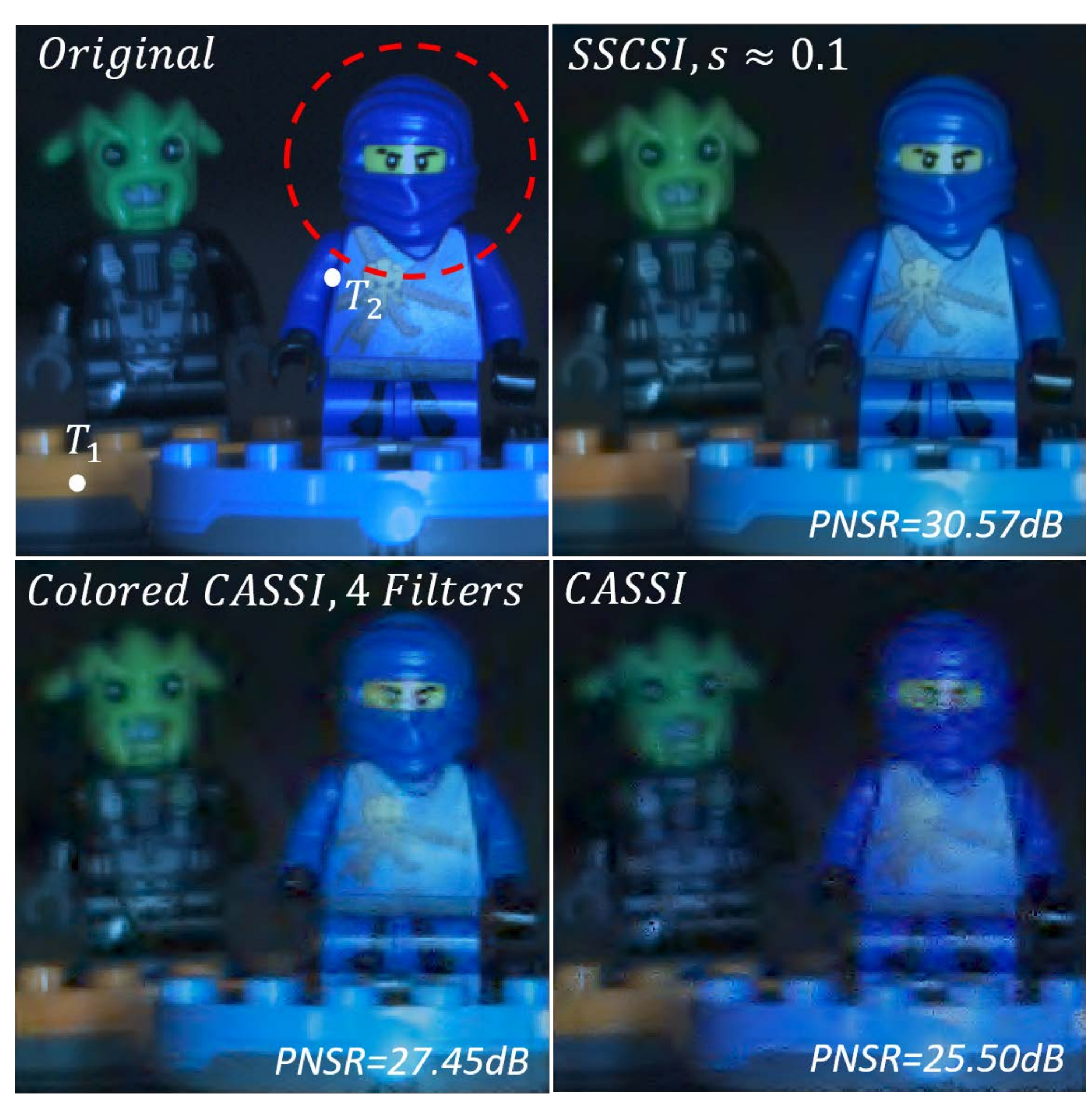}
	\caption{RGB profiles of the original and recovered datacubes. Top left: Original scene. Top right: SSCSI reconstruction, $s\approx0.1$. Bottom left: colored CASSI, 4 filters reconstruction. Bottom right: CASSI reconstruction. The simulations were done assuming $Q=3$ snapshots and $\Delta_c=\Delta_d$. The dashed red circle indicates the area to be zoomed in Fig. \ref{cassi_v_sscsi_3}.}
	\label{cassi_v_sscsi_2}
\end{figure}

\begin{figure}[h!]
	\centering
	\includegraphics[scale=0.3]{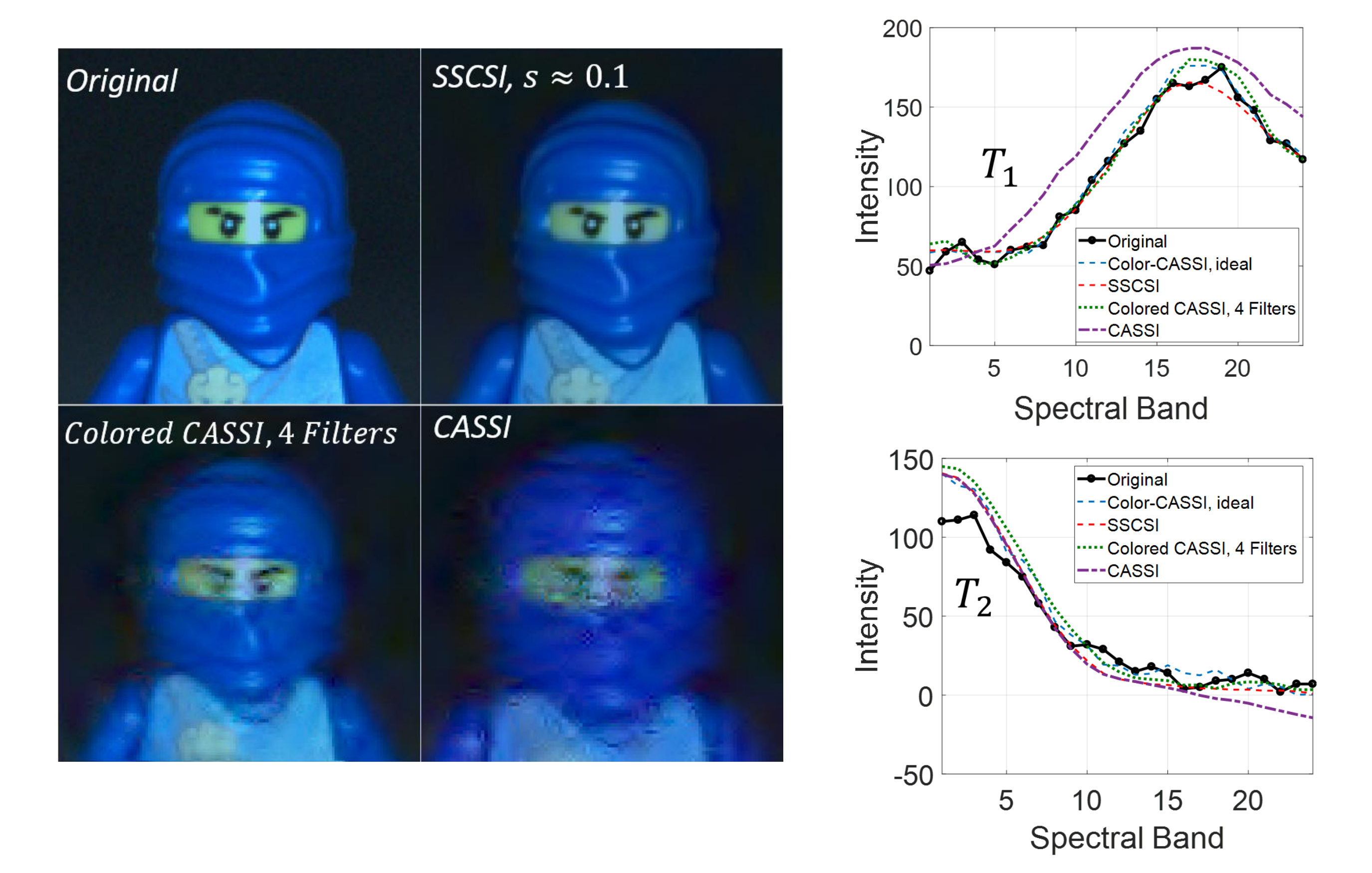}
	\caption{Left: zoomed portion of the RGB profiles indicated by the dashed red circle in Fig. \ref{cassi_v_sscsi_2}. Right: Spectral signatures for pixels $T_1$ and $T_2$. $T_1$ and $T_2$ are specified in Fig. \ref{cassi_v_sscsi_2}.}
	\label{cassi_v_sscsi_3}
\end{figure}

The recovery process of a hyperspectral scene with $L=24$ bands was simulated with a $N_d \times N_d$ detector array, with $N_d=256$, and $\Delta_d=\Delta_c$ using the CASSI, the colored CASSI and the SSCSI first order approximation developed in this paper. Since $\Delta_d=\Delta_c$, $\Delta_c/(1-s) > \Delta_d$ for any value of $s$. Therefore, the spatial size of the recovered datacube is going to be $ N_x \times N_y$, with $N_x=N_y=256$. Given that $L=24$ bands must be recovered, the coded aperture must be located at $s \approx 0.1$, according to Eq. (\ref{s_aprox}) and assuming that $\beta=1$. The fact that $\beta=1$ means that $\lambda_{min}$ impinges in the leftmost side of the coded aperture and $\lambda_{max}$ impinges on the rightmost side of the coded aperture, or $\alpha(\lambda_{max}-\lambda_{min})=N_c\Delta_c$.  The codes were generated such that the information captured in the shots is complementary, where the term complementarity is characterized  by the condition $\sum_{q=1}^{Q}\mathbf{t}^q_{m,n}=1 \ \forall \ m,n$;  this structure, called Boolean, has been proven to be optimal for the CASSI architecture \cite{bluenoise}; the transmittance of every coded aperture with  boolean structure is inversely proportional to the number of captured shots. The regularization parameter $\tau$ was empirically chosen to attain optimal results for all the scenarios, being in the majority of cases between $\tau=1e-5$ and $\tau=1e-4$. The Peak Signal-to-Noise Ratio (PSNR) was the parameter implemented in order to determine the quality of the reconstructions. The PSNR is defined as $20\log_{10}\left( \max_I/\mathrm{MSE}^{1/2}\right)$, where $\max_I$ is the maximum possible value of the image and $\mathrm{MSE}$ is the mean squared error with respect to the ground-truth. Figure \ref{cassi_v_sscsi} shows the PSNR  of the reconstructed scene as a function of the number of shots for the CASSI, the SSCSI first order approximation and the colored CASSI. The colored CASSI was implemented in two ways; in the first one, the coded aperture patterns were chosen such that the condition $\sum_{q=1}^{Q}\mathbf{t}^q_{m,n,k}=1 \ \forall \ m,n,k$ holds, with no limitation on the type of optical filters; this is called ideal colored CASSI. In the second one, the types of filters are limited to four: low pass, high pass, band pass and band stop, and the patterns on the mask were chosen such that the condition $\sum_{q=1}^{Q}\mathbf{t}^q_{m,n,k}\geq1 \ \forall \ m,n,k$ holds. Here, $\mathbf{t}^{q}_{m,n,k}$ represents the $k^{th}$ component of the spectral response of the optical filter located at position $(m,n)$ and shot $q$. As depicted, the SSCSI exhibits much better performance than both, the CASSI and the colored CASSI with 4 filters while achieving  similar results to the ideal colored CASSI. The comparison between the original scene and the recovered datacubes for $Q=3$ snapshots can be seen in Fig. \ref{cassi_v_sscsi_2}, while a zoomed version can be observed in Fig. \ref{cassi_v_sscsi_3} left. Here, the hyperspectral scenes were mapped to the RGB domain for visualization purposes. A comparison between the original and recovered signatures for two different pixels, can be seen in Fig. \ref{cassi_v_sscsi_3} right. As depicted, the SSCSI, the ideal colored CASSI and the colored CASSI with 4 filters show spectrally accurate results. 


\subsection{Spatial Super-Resolution}

\begin{figure}[h!]
	\centering
	\includegraphics[scale=0.42]{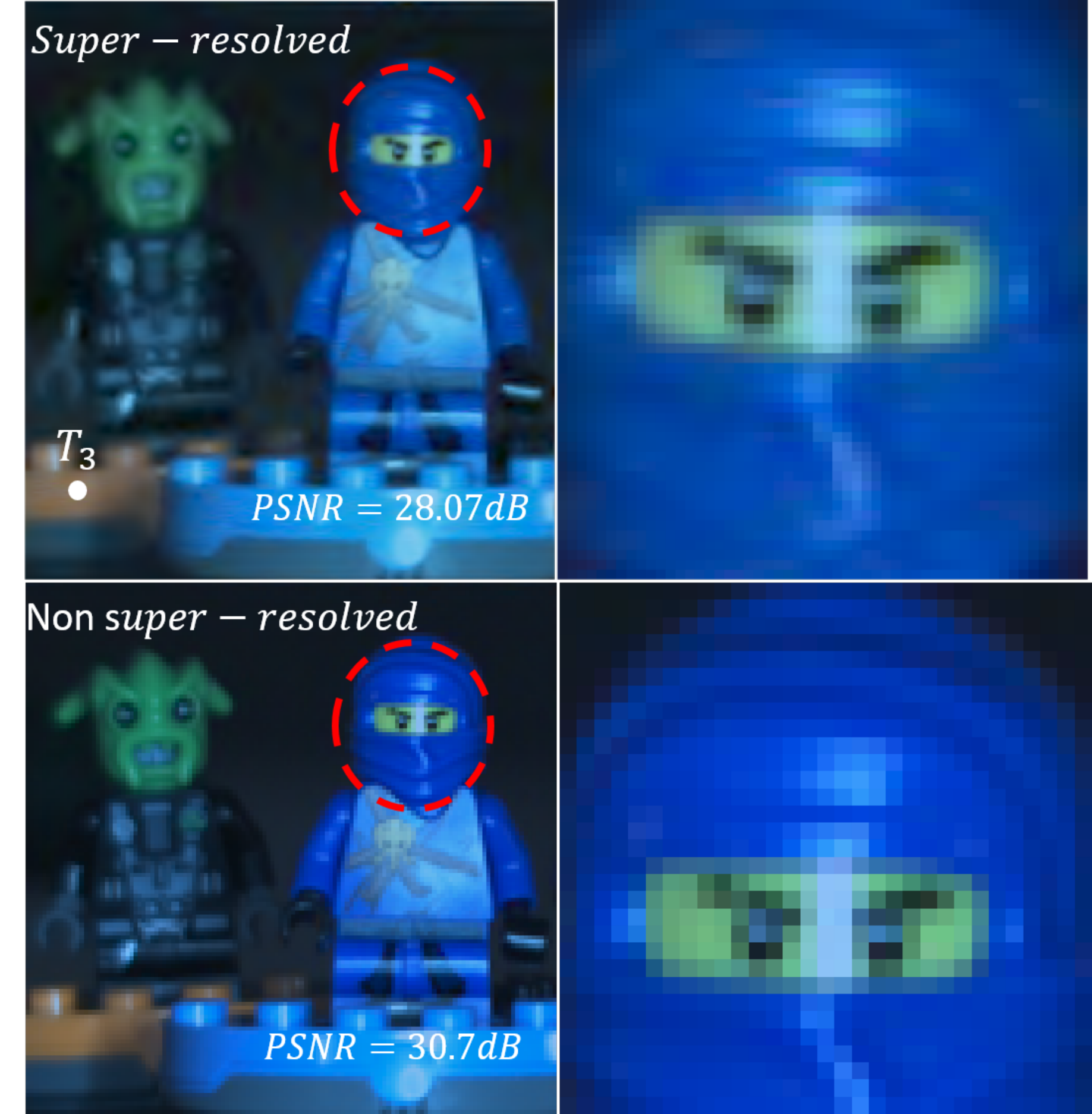}
	\caption{Top left: RGB profile of the super-resolved $233 \times 256 \times 24$ datacube, with $s\approx0.1$, $\Delta_c=\Delta_d/2$, and $CR=0.68$. Top right: Zoomed portion of the datacube. Bottom left: RGB profile of the  non super-resolved datacube, with $s\approx0.1$ and $\Delta_c=\Delta_d$. Bottom right: Zoomed portion of the datacube indicated by the dashed red circles.}
	\label{data_spat_3}
\end{figure}

\begin{figure}[h!]
	\centering
	\includegraphics[scale=0.45]{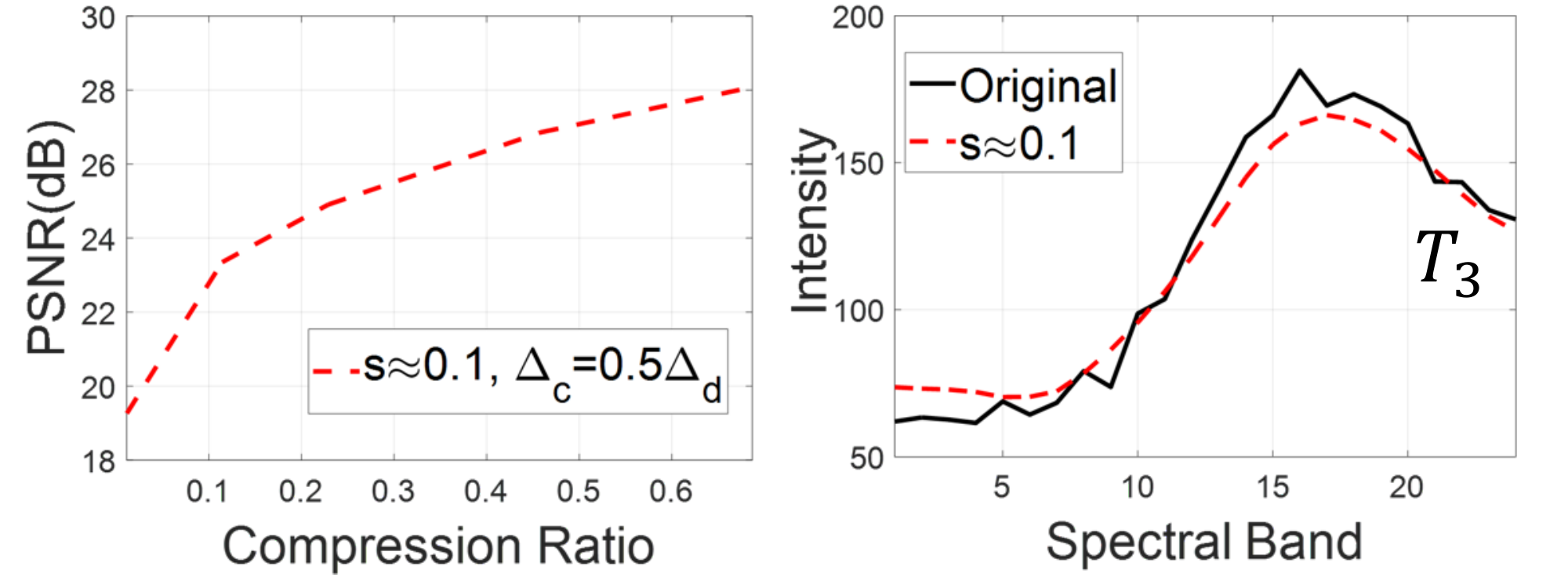}
	\caption{Left: PSNR of the super-resolved datacube as a function of the compression ratio, with $s\approx0.1$ and $\Delta_c=\Delta_d/2$. Right: Recovered spectral signature of $T_3$, for the super-resolved datacube and $CR=0.68$. $T_3$ is specified in Fig. \ref{data_spat_3}. This figure was implemented by the authors in \cite{Salazar_1}.}
	\label{data_spat}
\end{figure}

In \cite{2012spatial}, Arguello et al. proposed the reconstruction of spatially super-resolved hyperspectral scenes using CASSI, assuming that the ratio $\Delta_d/\Delta_c$ is an integer greater than 1. Likewise, with the SSCSI, Eq. (\ref{tot_7}) allows the reconstruction of spatially super-resolved datacubes with dimensions $\left \lceil N_d\frac{\Delta_d}{\Delta_c/(1-s)} \right \rceil \times N_d\frac{\Delta_d}{\Delta_c} \times L$. Consider $\Delta_c=\Delta_d/2$ and a $N_d \times N_d$ detector array, with $N_d=128$; with the SSCSI, a super-resolved datacube of spatial size $233 \times 256$ and $L=24$ spectral bands, can be recovered if  the coded aperture is located at $s\approx0.1$ and $\beta=1$. Figure \ref{data_spat_3} top left and right, shows the recovered super-resolved datacube when the compression ratio is $CR=0.68$. The compression ratio, $CR$, is defined as the ratio between the number of captured measurements and the size of the reconstructed datacube, or $CR=\frac{QN_d^2}{N_xN_yL}$.
Here, $N_d^2$ is the number of detector elements and $N_x \times N_y \times L$ is the size of the recovered datacube. The value of $CR$ must be less than $1$ in order to have compression during the capturing process. The $128 \times 128 \times 24$ non super-resolved datacube obtained from the same $128\times128$ detector array can be seen in Fig. \ref{data_spat_3} bottom left and right. A simple inspection of the target reveals the effect of the super-resolution on the spatial details of the reconstruction. The PSNR of the reconstructed $233 \times 256 \times 24$ super-resolved datacube as a function of the compression ratio can be seen in Fig. \ref{data_spat} left. Here, the compression ratio is incremented by increasing the number of captured snapshots $Q$.  A comparison between the original and the recovered spectral signatures can  be seen in Fig. \ref{data_spat} right.
\\

\vspace{-5mm}
\subsection{Spectral Zooming}
\begin{figure}[h!]
	\centering
	\includegraphics[scale=0.35]{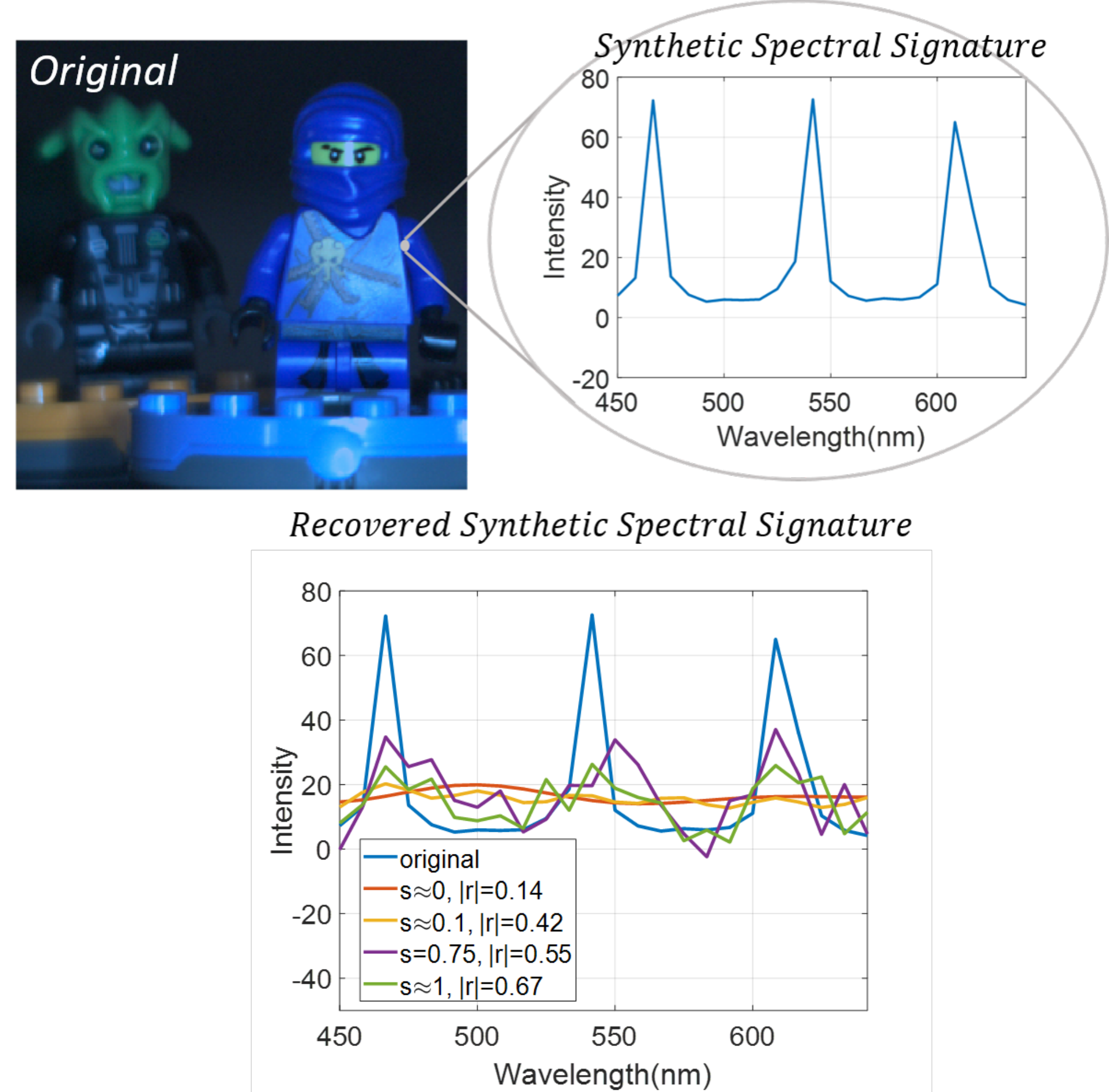}
	\caption{Top: Synthetic spectral signature created to evaluate the influence of the parameter $s$ on the spectral resolution. Bottom: Recovered synthetic spectral signature for different values of the parameter $s$ and a compression ratio of $CR=0.2$. The variable $|r|$ indicates the absolute value of the correlation coefficient between the original and the reconstructed signatures. This figure was previously implemented by the authors in \cite{Salazar_2}.}
	\label{syn}
\end{figure}

To illustrate the zooming process over the spectral dimension in a given scene, a spectral signature with strong peaks on three different wavelengths was introduced in the original datacube; the recovery was done for different values of $s$ and a compression ratio of $CR=0.2$. The conditions of $\beta=1$, $\Delta_d=\Delta_c$ and boolean coded apertures were assumed. As illustrated in Fig. \ref{syn}, no spectral details can be identified when the recovery is done at $s\approx0$ or $s\approx0.1$; when $s=0.75$ and $s\approx1$, on the other hand, the recovered spectral signatures are closer to the original one. This closeness is measured by the absolute value of the correlation coefficient $|r|$, between the original and reconstructed signatures.

\vspace{-3mm}
\subsection{Performance at different $s$}

\begin{figure}
	\centering
	\includegraphics[scale=0.68]{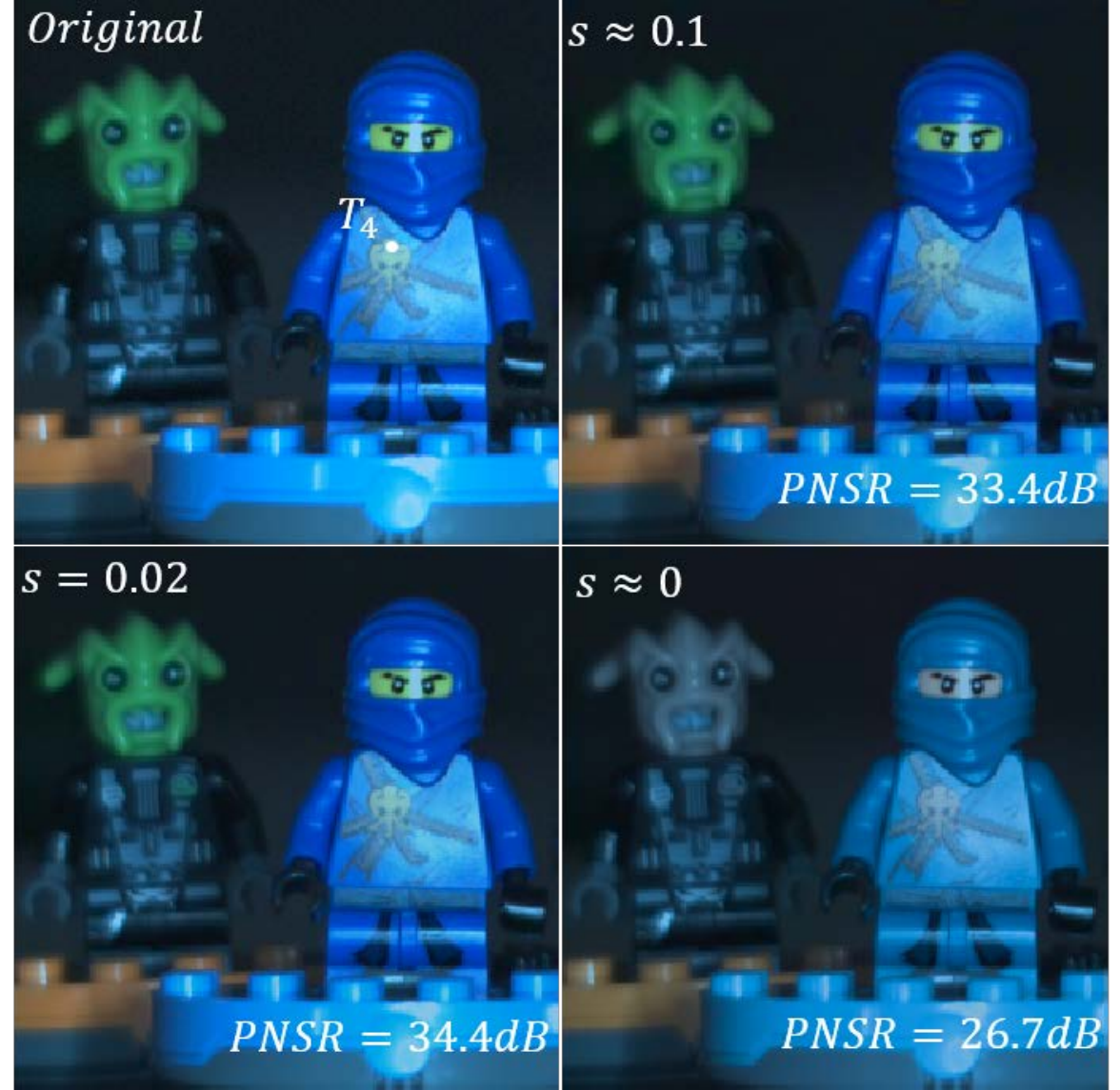}
	\caption{Original and reconstructed RGB profiles for 3 different coded aperture positions, $Q=6$ snapshots and $\Delta_c=\Delta_d$. The codes exhibit boolean structure. The optimal PSNR occurs at $s=0.02$, while at $s \approx 0$ the SSCSI does not recover spectral information.}
	\label{s_less_tog}
\end{figure}

\begin{figure}
	\centering
	\includegraphics[scale=0.3]{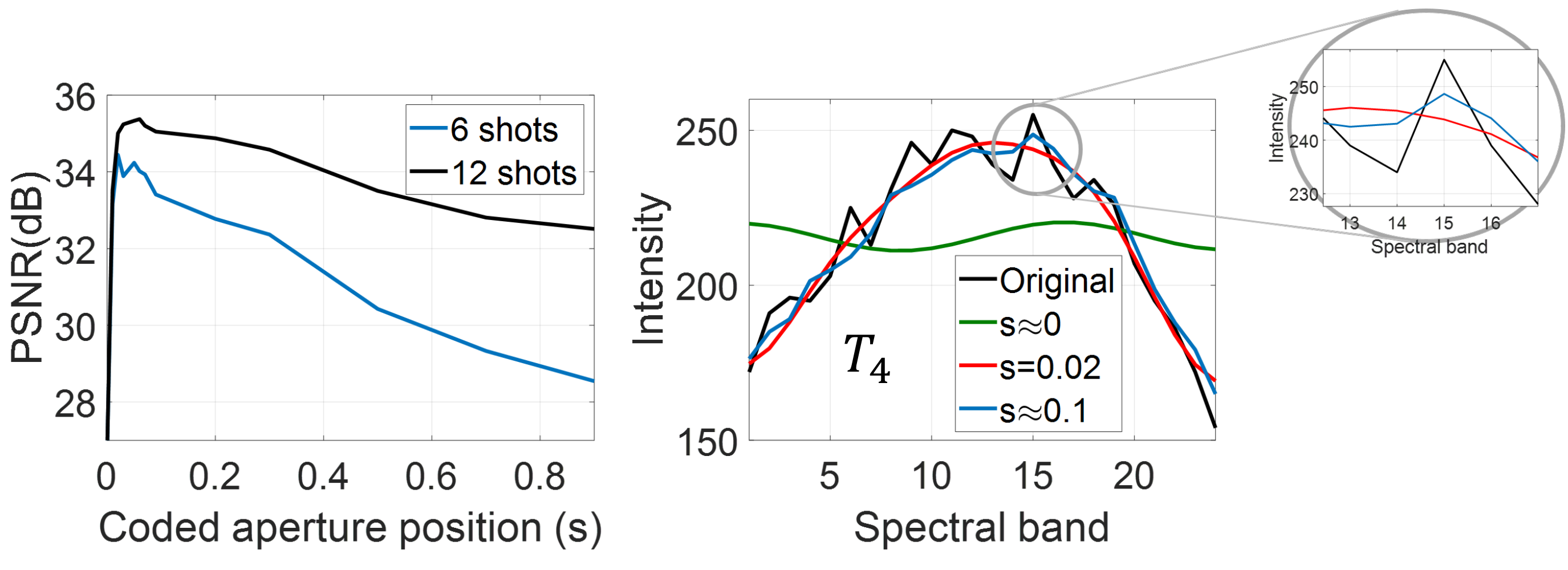}
	\caption{Left: PSNR of the reconstructed $256 \times 256 \times 24$ hyperspectral scene as a function of the coded aperture position $s$, for $Q=6$ and $Q=12$ snapshots. The codes exhibit boolean structure. Right: Recovered signature for pixel  $T_4$ and different values of $s$. The zoomed portion emphasizes the difference between the signatures and how the spectral details are lost for $s <0.1$. $T_4$ is specified in Fig. \ref{s_less_tog}. }
	\label{diff_s}
\end{figure}

As previously explained in section V-A, to recover a hyperspectral scene of $L=24$ bands using an $N_d \times N_d$ sensor, with $N_d=256$, and $\Delta_d=\Delta_c$, one must locate the coded aperture at $s\approx0.1$, if $\beta=1$. This subsection analyses the impact of recovering that same scene for  different values of $s$.  Figure \ref{s_less_tog} shows the RGB profiles of the recovered scene for three different values of $s$ and $Q=6$ snapshots. It is clearly seen that with $s\approx0$, the colors of the scene do not properly match the colors of the original datacube. 
Figure \ref{diff_s} left, shows the PSNR of the reconstructed $256 \times 256 \times 24$ hyperspectral scene as a function of the coded aperture position, for $Q=6$ and $Q=12$ snapshots. As depicted, if $s\approx0$, the quality of the reconstruction is low, since the system is not able to distinguish any spectral information of the scene. Increasing $s$ leads to enhance the quality of the reconstruction. However, when $s$ gets close to $1$, the PSNR drops as it was also shown in \cite{lin}. The comparison between the spectral signatures for different values of $s$ can be seen in Fig. \ref{diff_s} right; reducing $s$ to $0.02$ makes the fine details of the curve to be lost while maintaining its shape; for $s\approx0$, no spectral information is recovered.



\subsection{Coherence of the sensing matrix}

In compressive sensing, the coherence of the sensing matrix $\mu(\mathbf{A})$, where $\mathbf{A}=\mathbf{H}\boldsymbol{\Psi}$, has been extensively used as a measure of the quality of the sensing process, where matrices with a low coherence lead to high quality reconstructions~\cite{elad,paradacolor,duarte,candes2007}.  Table I contains the PSNR values of the reconstructions and the coherence of the sensing matrix $\mu(\mathbf{A})$ for different  values of $s$. Here, a $N_d \times N_d$ sensor array with $N_d=64$ was considered,  with $\Delta_d=\Delta_c$ and $\beta=1$.
A multispectral scene of dimensions  $64\times64\times 8$ was recovered and the coded apertures exhibit boolean structure. As it can be appreciated in Table I, $\mu(\mathbf{A})$ is connected to the quality of the recovered scene and low values of $\mu(\mathbf{A})$ are associated to high PSNR values. Further analysis of the coherence in SSCSI will be done in future work.
\begin{table}[h!]
	\centering
	\caption{Coherence of the sensing matrix for different values of $s$}
	\begin{tabular}{l c c}
		\toprule[1.5pt]
		s &PSNR(dB) & $\mu(\mathbf{A})$  \\
		\hline
		0 & 26.7 & 1\\ 
		0.01 & 28.53 & 0.997\\
		0.02 &30.22 &0.971\\
		0.03 &30.35 &0.971\\
		0.05 &30.71 &0.922\\
		0.07 &31.54 &0.888\\
		\toprule[1.5pt]
	\end{tabular}
	\label{mutual}
\end{table} 

\section{Experimental measurements}

\begin{figure}[h!]
	\centering
	\includegraphics[scale=0.5]{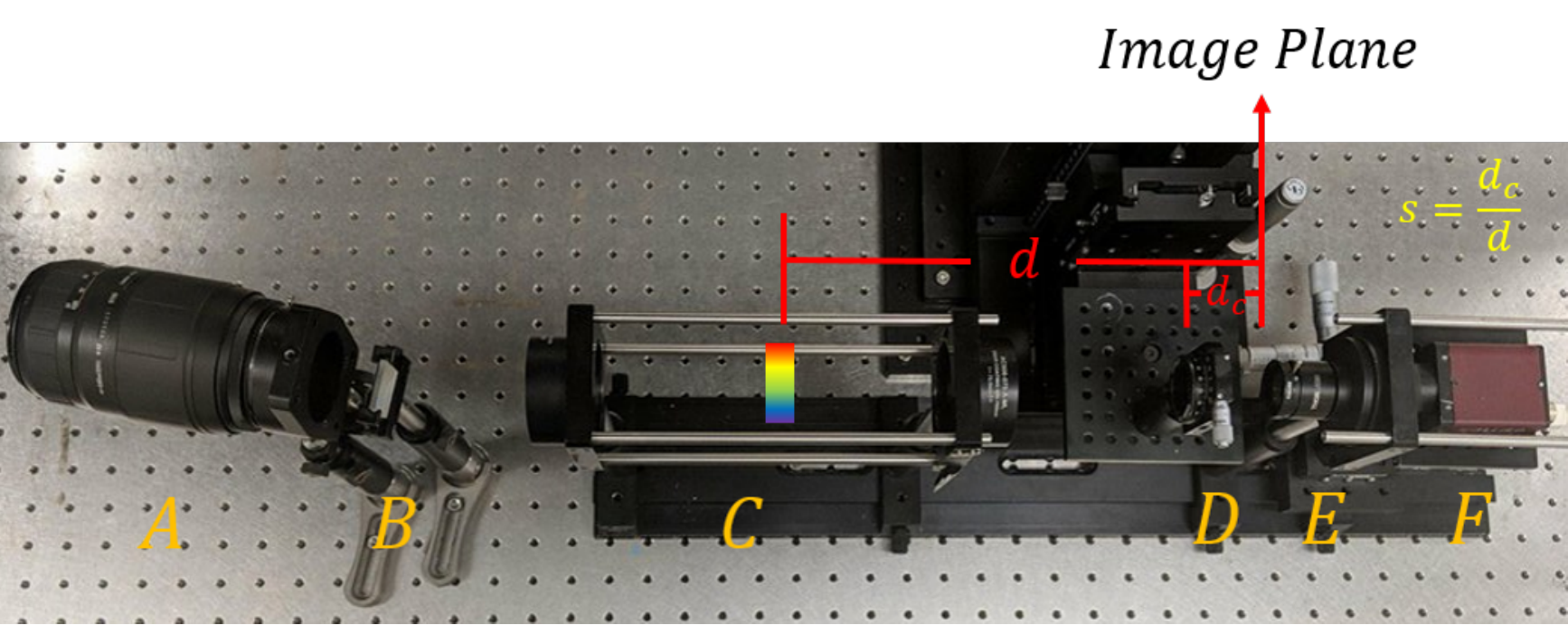}
	\caption{SSCSI experimental set up. 6 main components are distinguished. A. TAMRON AF 70-300mm objective lens. B. 300 grooves/mm transmissive diffraction grating. C. 4f system composed of two 75mm, 2'' lenses. D. Coded aperture. E. Relay system composed of a 35mm, 1'' lens. F. Stingray\textsuperscript{TM} $640 \times 480$ CCD monochrome camera with 9.9\textmu m pitch size. As depicted, the 4f system allows to have more flexibility when displacing the coded aperture.  Notice that the optical arm is bended according to the diffraction grating angle.}
	\label{set_exp}
\end{figure}

The SSCSI was experimentally implemented as depicted in Fig. \ref{set_exp}.  A TAMROM AF 70-300mm lens locates an image of the scene at the transmissive diffraction grating (300 grooves/mm). The diffracted image is then reshaped using a 4f system composed of two 75mm, 2'' lenses. In a 4f system, the image located at a distance $f$ with respect to the first lens is reshaped at a distance $f$ with respect to the second lens. The 4f system provides major flexibility in the movement of the coded aperture, in the sense that $s=0$ is accessible, since it is located in a spatial position far from the sensor. The spectral plane will be located in between the lenses that compose the 4f system. The coded scene is then reshaped at the sensor by a 1'', 35mm lens. The implemented sensor is a Stingray\textsuperscript{TM} $640 \times 480$ CCD monochrome camera with 9.9\textmu m  pitch size. The spectral range of interest was defined as 480nm-620nm.

\vspace{-4mm}
\subsection{Spectral Resolution}

\begin{figure}[H]
	\centering
	\includegraphics[scale=0.28]{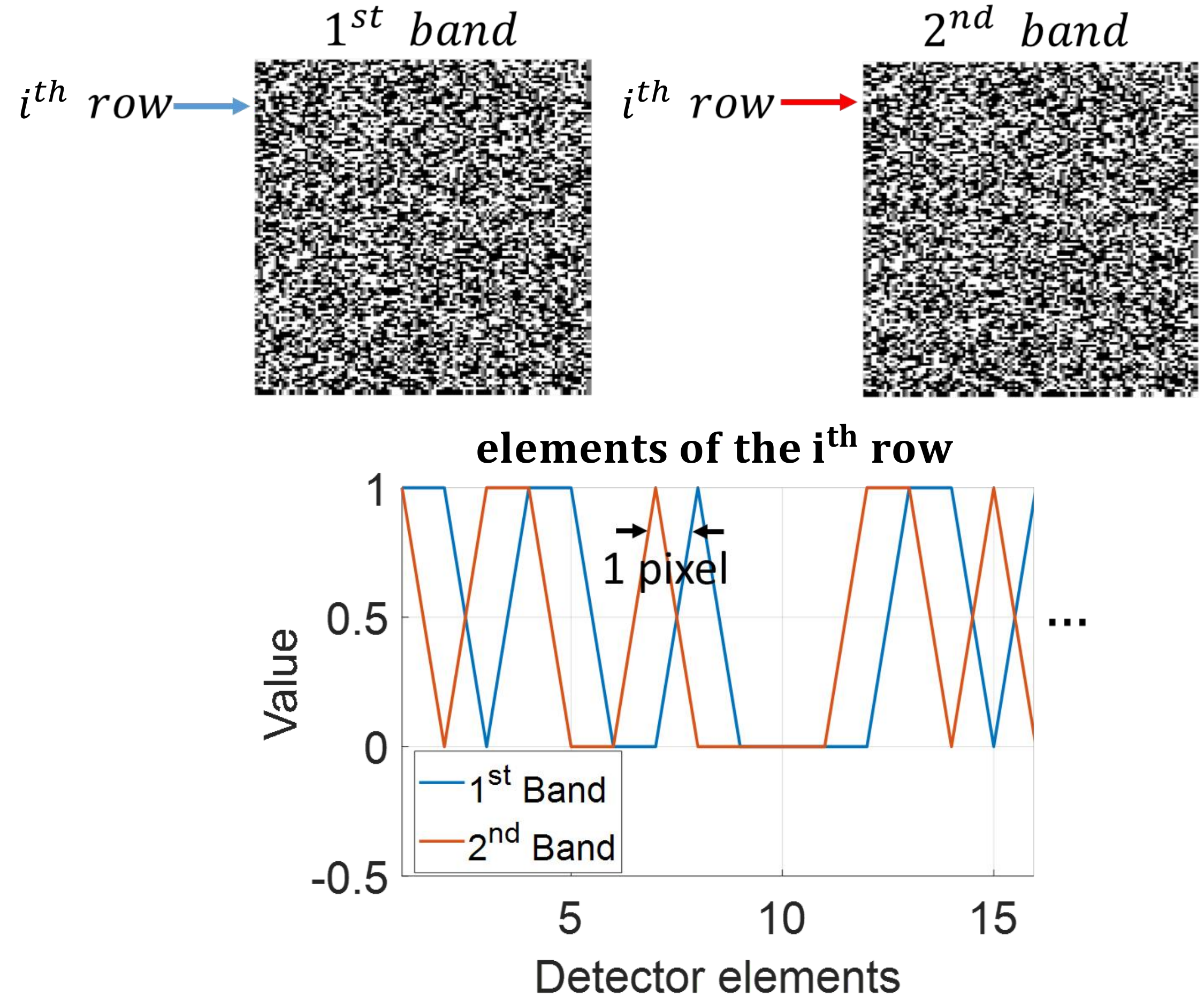}
	\caption{Experimental calculation of the spectral resolution. Top: effective coded aperture for two consecutive bands. Bottom: Comparison of  $i^{th}$ effective coded aperture row elements for the two consecutive bands. In order for the bands to be resolvable, the curves must be shifted by one pixel at the sensor. }
	\label{spec_criteria}
\end{figure}

The theoretical spectral resolution can be determined by defining the size of the spectral plane with respect to the coded aperture ($\beta$ in (\ref{s_aprox})). It was found that the 480nm-620nm spectral range occupies a physical space three times the coded aperture width, therefore, $\beta\approx3$. This was done by locating a white board on the spectral plane and measuring the physical distance between $\lambda_{min}=480$nm and $\lambda_{max}=620$nm. The number of resolvable bands with $\Delta_c/\Delta_d=1$ and $N_d=N_c=256$ is given by $L=\left \lceil 256s\beta \right \rceil$. On the other hand, the criteria used to determine whether two adjacent spectral bands are resolvable in experiments, can be seen in Fig. \ref{spec_criteria}. As depicted, the patterns that code two adjacent spectral bands at a given $s$ are first captured with the sensor; then, the $i^{th}$ rows from those patterns are compared. In order for the two bands to be resolvable, the curves must be shifted by one pixel. Table II shows the experimental and theoretical spectral resolution for different values of $s$.

\begin{table}[H]
	\centering
	\caption{Theoretical and experimental spectral resolution for different $s$ parameters}
	\begin{tabular}{l c c}
		\toprule[1.5pt]
		s &$\Delta_{\lambda}$, Theoretical (nm) & $\Delta_{\lambda}$, Experimental (nm)\\
		\hline
		0.004 & 43 & 40\\
		0.0078 &22 &28\\
		0.011 &16 &20\\
		\toprule[1.5pt]
	\end{tabular}
	\label{exp_resol}
\end{table} 

%
%
 \subsection{Experimental Datacube Reconstruction}
 
\begin{figure}
	\centering
	\includegraphics[scale=0.16]{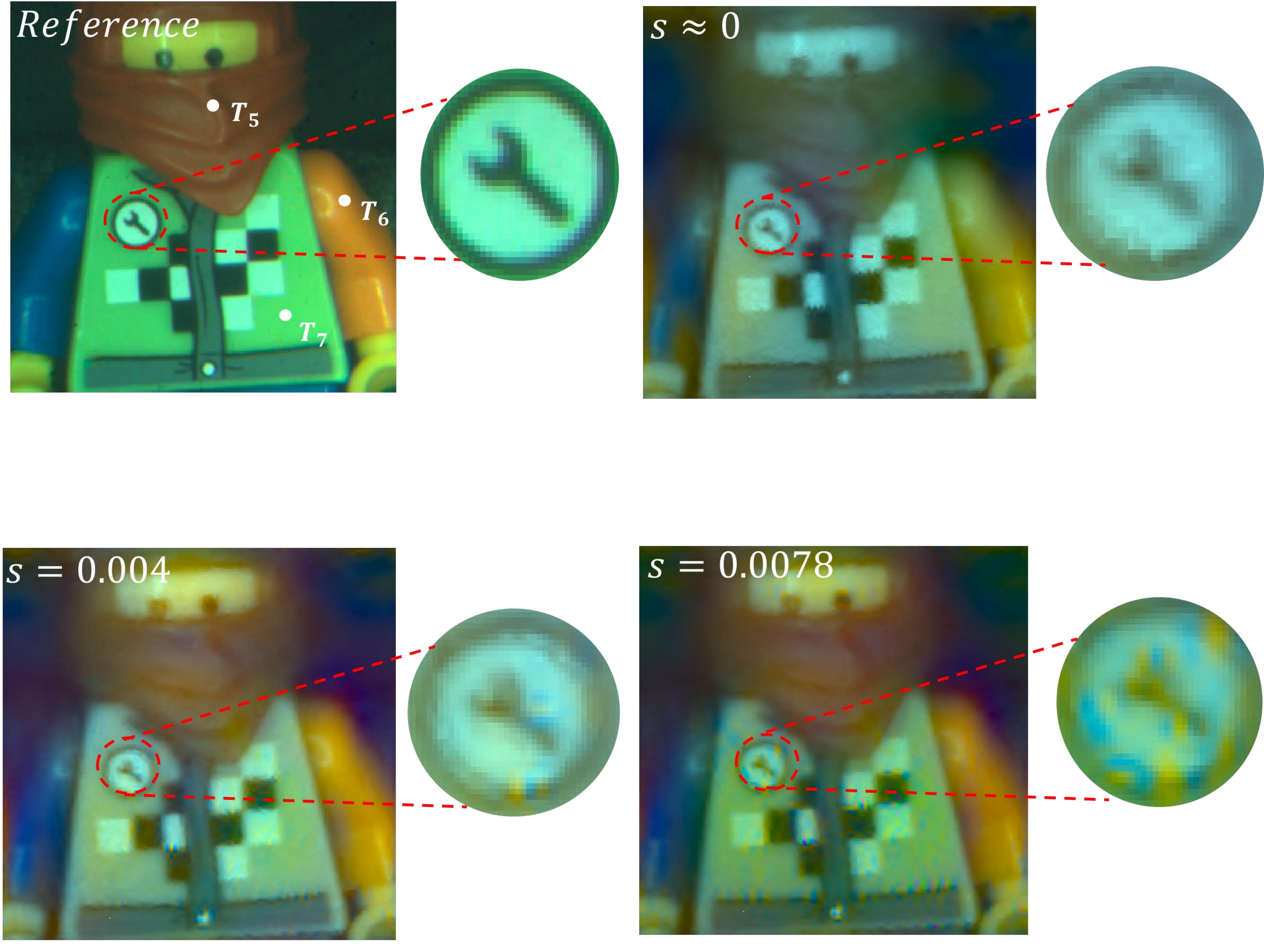}
	\caption{RGB profiles of the reference scene and the reconstructed datacubes for $3$ different values of $s$ and two complementary shots. $s \approx 0$, $s=0.004$, and $s=0.0078$. Increasing $s$ allows the colors of the hyperspectral scene to be better distinguished and, at the same time the spatial quality drops as it can be seen in the zooming portion of the scenes.}
	\label{2_shots}
\end{figure}

\begin{figure}
		\centering
	\includegraphics[scale=0.11]{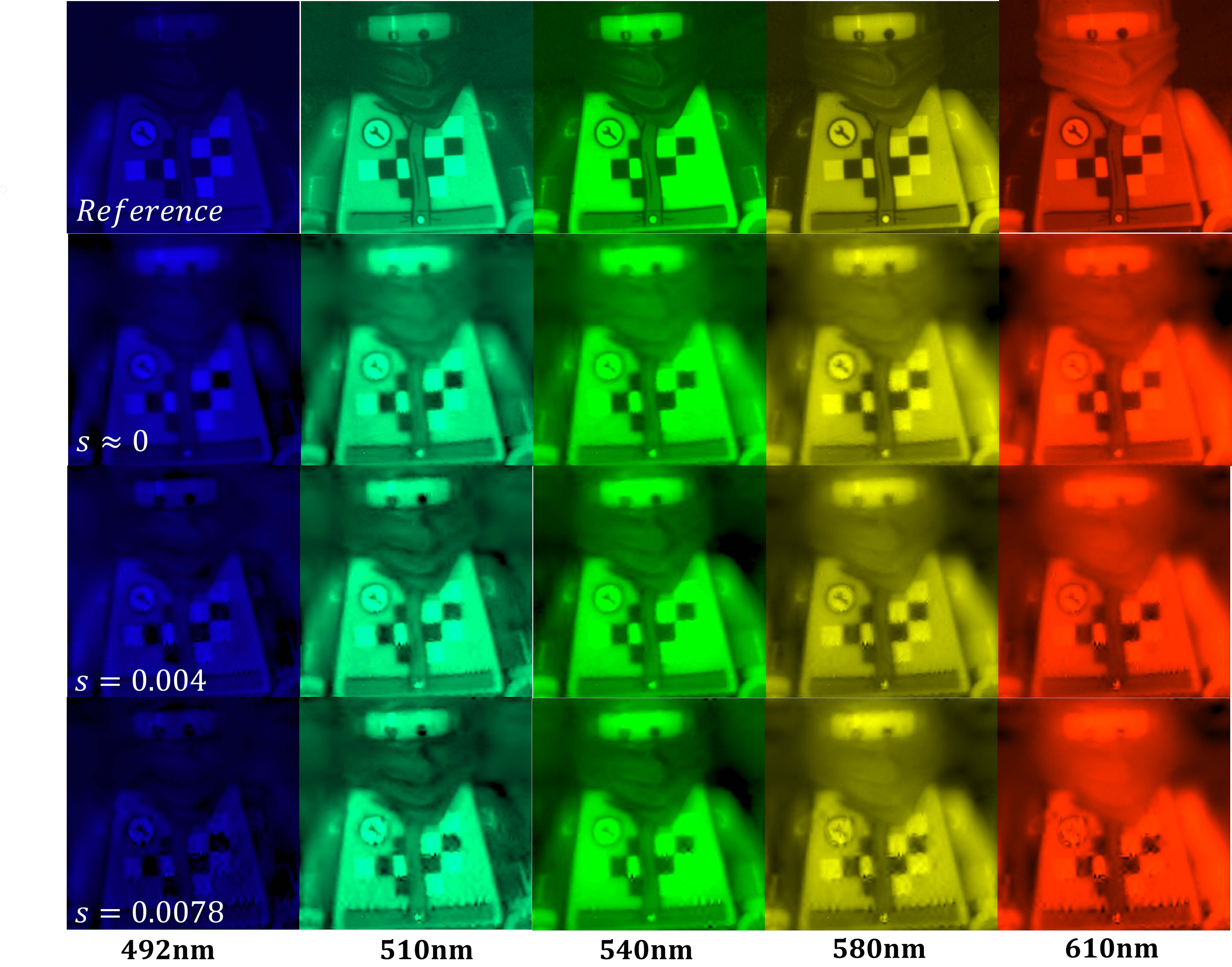}
	\caption{Reference and reconstructed spectral bands for $3$ different values of $s$ and two complementary shots. From top to bottom: Reference, $s \approx 0$, $s=0.004$, and $s=0.0078$.}
	\label{2_shots_2}
\end{figure}

\begin{figure}
	\centering
	\includegraphics[scale=0.35]{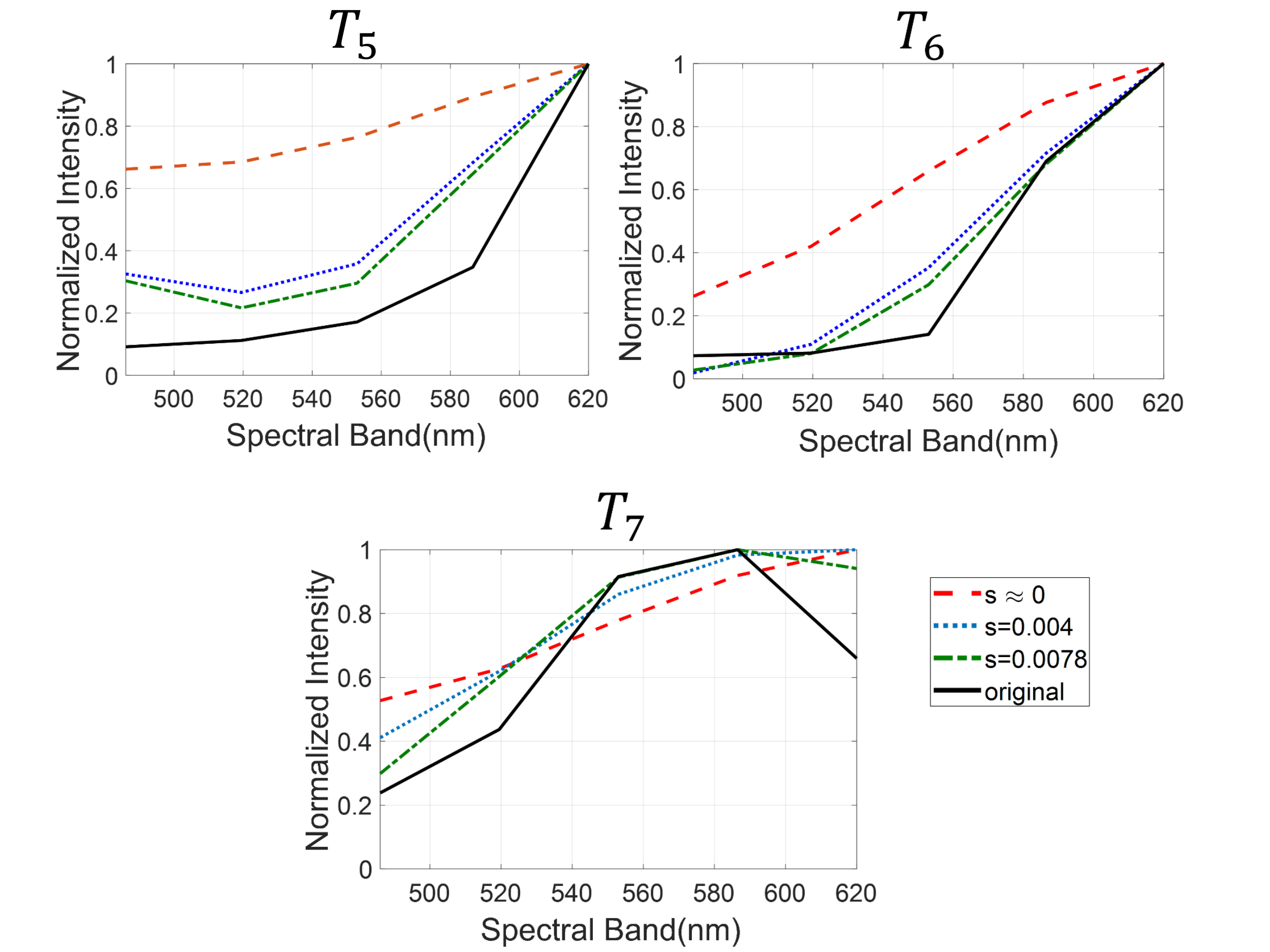}
	\caption{Original and reconstructed spectral signatures for pixels $T_5$, $T_6$, and $T_7$ and $3$ different values of $s$, $s \approx 0$, $s=0.004$, and $s=0.0078$. $T_5$, $T_6$, and $T_7$ are specified in Fig. \ref{2_shots}.}
	\label{spec_exp}
\end{figure}

A hyperspectral scene was reconstructed using $2$ snapshots and different values of $s$. The coded aperture patterns exhibit boolean structure and were printed on a photo-mask with pitch size of $\Delta_c=19.4$\textmu m. The matrix $\mathbf{H}$ was generated by first locating a white board target and then capturing the coded aperture pattern for each spectral band to be recovered, at a given $s$. Then, that pattern is located in the respective diagonal of the matrix $\mathbf{H}$; this is done for every captured snapshot. After that, the white board is replaced by the hyperspectral scene, and the snapshots are properly captured. The reconstruction algorithm is posteriorly executed to find the optimal coefficients $\boldsymbol{\pi}$. The RGB profiles of the reference scene and the reconstructed datacubes can be seen in Fig. \ref{2_shots}, while the spectral bands and recovered spectral signatures are contained in Fig. \ref{2_shots_2} and Fig. \ref{spec_exp}. As depicted, increasing $s$  allows the spectral information of the scene to be better distinguished.

%


The SSCSI bases its functionality on the location of the coded aperture in an out of focus position or $s>0$, as depicted in Fig. \ref{set_up_2}. As mentioned in Section II, the present model assumes an infinitesimally small aperture in the objective lens. However, in experiments, the finite size of the aperture introduces blurring effects when $s>0$ and, in fact, increasing $s$  deteriorates the spatial quality of the reconstructed hyperspectral scene. Figure \ref{2_shots} shows the zooming over one portion of the recovered scene for the different values of $s$. It is evident that the colors of the scene become more noticeable for a bigger $s$, but at the same time, the spatial quality drops.

There are several factors to consider in the experiments. Besides the blurring proportional to $s$, that affects the spatial quality of the reconstructions, one might find other challenges when implementing the SSCSI. One of them is the amount of light that passes through the system, since the aperture of the objective lens must be as small as possible. At the same time, the size of this aperture directly affects the resolution of the SSCSI, given that the minimum  resolvable feature for any optical system is determined by the diffraction limit. On the other hand, the fact that the spectral plane is three times bigger than the coded aperture  $(\beta\approx3)$, leads to  extreme wavelengths to impinge out of the mask; this means, in practice, that the range of wavelengths is reduced to the ones impinging inside the mask.

\section{Conclusions and further steps}
This papers develops a rigorous discretization measurement model of the Spatial Spectral Compressive Spectral Imager (SSCSI), based on the physical dimensions of the coded aperture and the detector array, and the spectral dispersion introduced by the diffraction grating, characterized by $\alpha$. Two different scenarios were proposed:

\begin{itemize}
	\item $\Delta_d \leq \Delta_c/(1-s)$: This scenario is modeled by Eq. (\ref{tot_7}), where $W_{m,m'}$ is used to model the mismatch between the coded aperture and the detector array. The spatial resolution is defined as $\Delta_c/(1-s)\times\Delta_c$. As described in Section V-B, a spatially super-resolved datacube of dimensions$\left \lceil N_d\frac{\Delta_d}{\Delta_c/(1-s)} \right \rceil \times N_d\frac{\Delta_d}{\Delta_c}\times L$,  can be recovered when this condition holds.
	\item $\Delta_d > \Delta_c/(1-s)$: This scenario is modeled by Eq. (\ref{disc_10}), where an effective mask is generated by the combination of coded aperture pixels proportional to the areas, using the variable $p_m$.  The spatial resolution is defined as $\Delta_d\times\Delta_c$
\end{itemize}

The number of spectral bands that can be recovered using the SSCSI was found to depend on three variables:  The coded aperture pitch size $\Delta_c$, the spectral dispersion introduced by the diffraction grating characterized by $\alpha$, and the coded aperture position represented by the parameter $s$. In fact, it was found that a bigger $s$ increases the number of resolvable spectral bands. Therefore, a movement of the coded aperture towards the spectral plane can be seen as performing a spectral zooming over a given hyperspectral scene.  The SSCSI performance, in experiments, depends on the aperture of the objective lens. Increasing the size of the aperture makes the spatial quality of the recovered scene to drop as the coded aperture is displaced towards the spectral plane. On the other hand, reducing the aperture also decreases the amount of light passing through the system.\\
\indent Ongoing work includes the optimization of the coded aperture patterns for SSCSI based on the concept of coherence and the proposed model in this paper. Similarly, a more accurate discretization model is being analyzed.

\section*{Acknowledgment}

This research project was funded by the Department of Homeland Security, Science and Technology Directorate (Contract $\mathrm{\#}$HSHQDC-15-C-B0016).
\section*{Appendix A\\ $W_{m,m'}$ and $\tilde{\mathbf{t}}_{m,n'}$ full calculation}

\subsubsection*{$W_{m,m'}$}
\begin{figure}[H]
	\centering
	\includegraphics[scale=0.5]{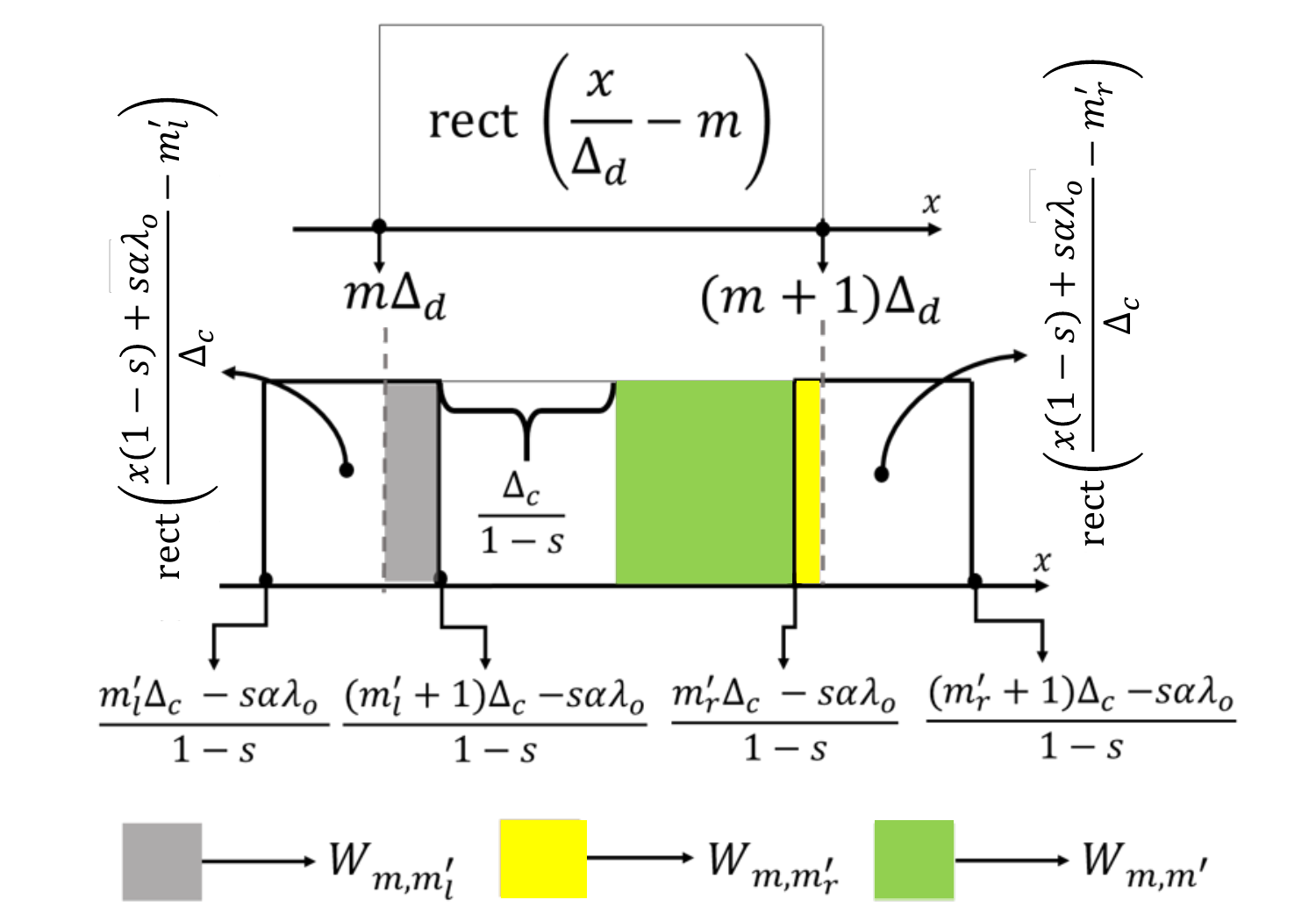}
	\caption{Graphical explanation to calculate $W$ parameters implemented in Eq. \ref{disc_5}.}
	\label{W_Calc}
\end{figure}

As mentioned in Section III-A, $W_{m,m'}$ indicates the fraction of the coded aperture $\mathbf{t}_{m',n'}$ impinging on the sensor element $(m,n)$. A graphical description of $W_{m,m'}$ can be seen in Fig. \ref{W_Calc}. 
As depicted, three cases can be distinguished; when $m'=m'_l$ (gray area in Fig. \ref{W_Calc}), $W_{m,m'_l}$ can be calculated as $W_{m,m'_l}=\frac{\frac{(m'_l+1)\Delta_c-s\alpha\lambda_o}{1-s}-(m)\Delta_d}{\Delta_c/(1-s)}$. The term $\Delta_c/(1-s)$ is used as a normalization parameter that allows one to obtain the percentage of $\mathbf{t}_{m'_l,n'}$ measured by the sensor element $(m,n)$. Likewise, when  $m=m'_r$ (yellow region in Fig. \ref{W_Calc}), $W_{m,m_r'}$ can be calculated as $W_{m,m'_r}=\frac{(m+1)\Delta_d- \frac{m'_r\Delta_c-s\alpha\lambda_o}{1-s}}{\Delta_c/(1-s)}$. Finally, if $m'_l<m'<m'_r$ (green area in Fig. \ref{W_Calc}), $W_{m,m'}$ can be calculated as $W_{m,m'}=\frac{\Delta_c/(1-s)}{\Delta_c/(1-s)}=1$. The fact $W_{m,m'}=1$ indicates that $\mathbf{t}_{m',n'}$ is fully impinging on sensor element $(m,n)$. Equation (\ref{W_1}) contains a general expression for $W_{m,m'}$ considering all the mentioned cases.
An example of how to calculate $W_{m,m'}$ can be done based on Fig. \ref{gen_mod} right. Here, $W_{0,2}=\frac{\Delta_d-\frac{2\Delta_c}{(1-s)}}{\Delta_c/(1-s)}$, $W_{1,2}=\frac{\frac{3\Delta_c}{(1-s)}-\Delta_d}{\Delta_c/(1-s)}$ and $W_{0,1}=1$.\\

\subsubsection*{$\tilde{\mathbf{t}}_{m,n'}$}
\begin{figure}[h!]
	\centering
	\includegraphics[scale=0.3]{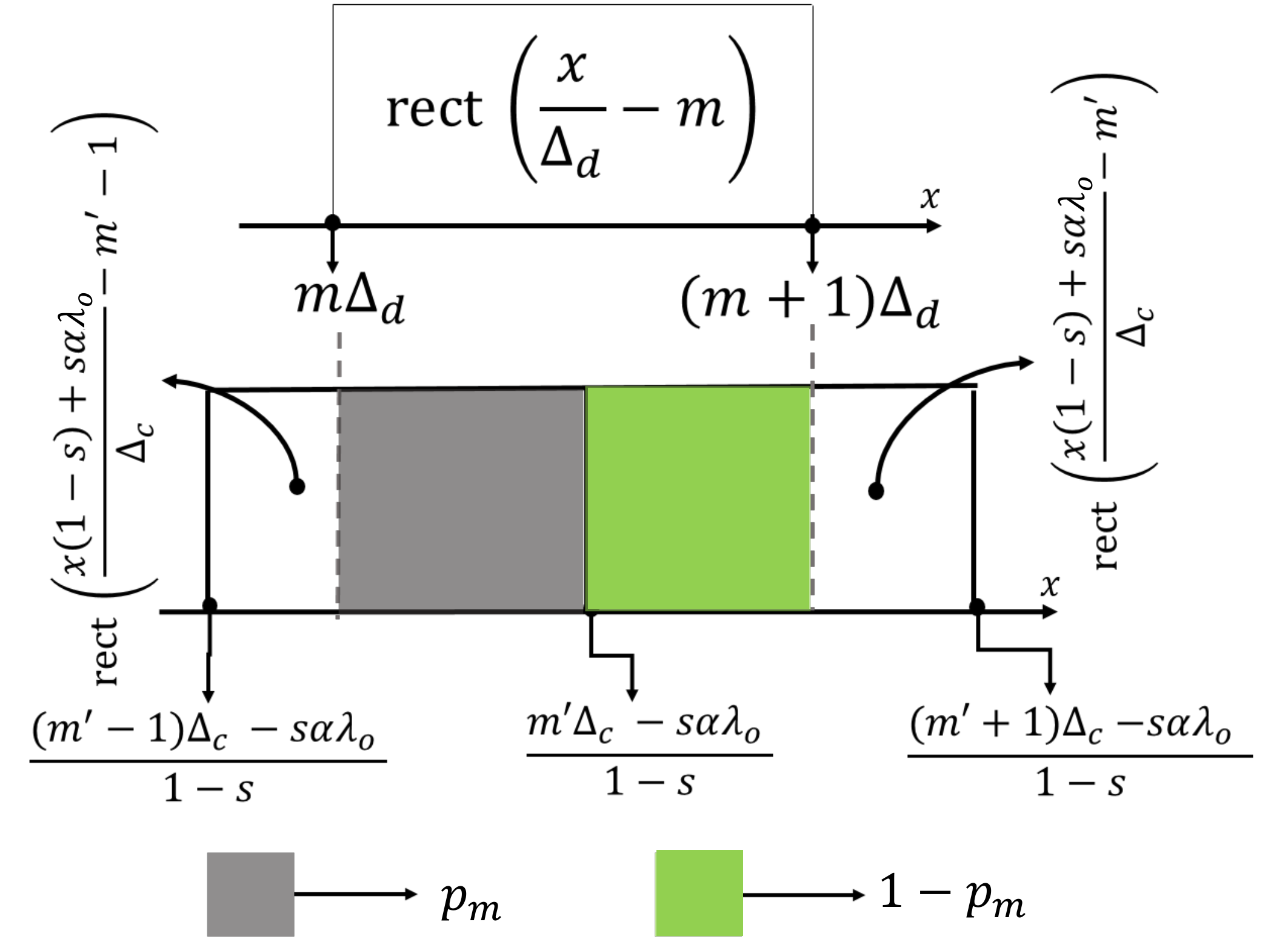}
	\caption{Graphical explanation to calculate the variable $p_m$ in Eq. \ref{disc_9}.}
	\label{annex_t}
\end{figure}

The effective coded aperture $\tilde{\mathbf{t}}_{m,n'}$ is given by the expression $\tilde{\mathbf{t}}_{m,n'}=\mathbf{t}_{m'-1,n'}\times p_m + \mathbf{t}_{m',n'}\times (1-p_m)$, 
where $n'$ is given by Eq. (\ref{eq_n}), $m'=\left \lfloor \frac{(m)\Delta_d(1-s)+s\alpha\lambda_o}{\Delta_c} \right \rfloor +1$ and the parameter $p_m$ indicates the percentage of the $(m,n)^{th}$ sensor element occupied by the $(m'-1,n')^{th}$ coded aperture element. Figure \ref{annex_t} contains a graphical explanation to calculate $p_m$. Notice that, according to this figure,  $m\Delta_d \leq \frac{m'\Delta_c-s\alpha\lambda_o}{1-s}\leq(m+1)\Delta_d$, or equivalently $m'=\left \lfloor \frac{(m)\Delta_d(1-s)+s\alpha\lambda_o}{\Delta_c} \right \rfloor +1$ (considering that $m'$ is an integer index). 

The portion of the $(m,n)^{th}$ sensor element occupied by  $\mathbf{t}_{m'-1,n'}$ (gray area in Fig. \ref{annex_t}) can be calculated as $p_m=\frac{\frac{m'\Delta_c-s\alpha\lambda_o}{(1-s)}-(m)\Delta_d}{\Delta_d}$, where the sensor pitch size $\Delta_d$ appears in the denominator for normalization purposes. If $\frac{m'\Delta_c-s\alpha\lambda_o}{1-s}\geq(m+1)\Delta_d$,  $\mathbf{t}_{m'-1,n'}$ fully occupies the $(m,n)^{th}$ sensor element  and therefore $p_m=1$. A general expression to calculate $p_m$, including the two cases mentioned can be seen in Eq. (\ref{p_2}).
\vspace{-2mm} 
\section*{Appendix B\\ Spectral Resolution calculation}

\subsubsection*{$\frac{\Delta_c}{1-s} \leq \Delta_d$}

\begin{figure}[H]
	\centering
	\includegraphics[scale=0.3]{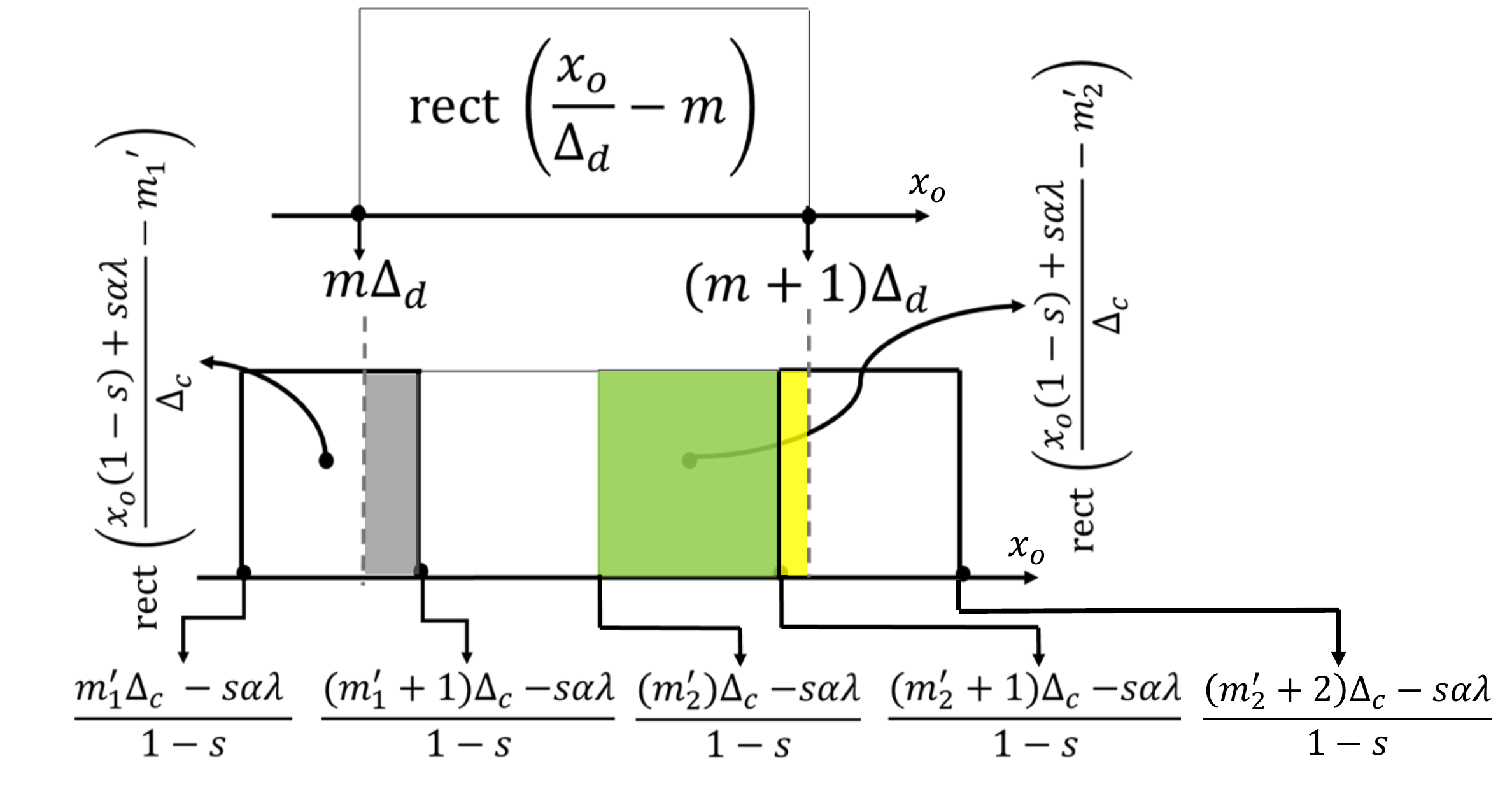}
	\caption{Graphical explanation to calculate the spectral resolution of the SSCSI, when $\frac{\Delta_c}{1-s} \leq \Delta_d$.}
	\label{ANNEX_B}
\end{figure}

The spectral resolution of SSCSI is given by the length of the region where Eq. (\ref{spec_2}) is different from zero. This is defined by the overlapping of the two rectangular functions in that expression, which can be seen in Fig. \ref{ANNEX_B}. As depicted two concrete cases can be analyzed. In the first one (gray area in Fig. \ref{ANNEX_B}), $m\Delta_d \leq \frac{(m'_1+1)\Delta_c-s\alpha\lambda}{1-s}$ and $m\Delta_d \geq \frac{m'_1\Delta_c-s\alpha\lambda}{1-s}$ or, in other words, $\frac{m'_1\Delta_c-m\Delta_d(1-s)}{s\alpha} \leq \lambda \leq  \frac{(m'_1+1)\Delta_c-m\Delta_d(1-s)}{s\alpha}$. This interval has an extension of $\Delta_{\lambda}=\frac{\Delta_c}{s\alpha}$. Notice that the same $\Delta_{\lambda}$ is obtained for the yellow region in Fig. \ref{ANNEX_B}. 

 In the second case (green area), $m\Delta_d \leq \frac{m'_2\Delta_c-s\alpha\lambda}{1-s}$ and $(m+1)\Delta_d \geq \frac{(m'_2+1)\Delta_c-s\alpha\lambda}{1-s}$, or equivalently $\frac{(m'_2+1)\Delta_c-(m+1)\Delta_d(1-s)}{s\alpha} \leq \lambda \leq  \frac{m'_2\Delta_c-m\Delta_d(1-s)}{s\alpha}$, which has an extension of $\Delta_{\lambda}=\frac{\Delta_d(1-s)-\Delta_c}{s\alpha}$.\\
  Let the ratio $\Delta_d/\Delta_c=C$, where $C$ is an integer greater than 1. In order to make $\frac{\Delta_c}{s\alpha}> \frac{\Delta_d(1-s)-\Delta_c}{s\alpha}$, one can prove that $s > 1-\frac{2}{C}$. For $C=2$ (the case analyzed in section V-B),  the inequality $\frac{\Delta_c}{s\alpha}> \frac{\Delta_d(1-s)-\Delta_c}{s\alpha}$ holds for $s>0$. Therefore, $\Delta_{\lambda}=\frac{\Delta_c}{s\alpha}$ can be taken as an upper bound of the spectral resolution. When $C>2$, $\frac{\Delta_c}{s\alpha}< \frac{\Delta_d(1-s)-\Delta_c}{s\alpha}$ for certain values of $s$, and $\Delta_{\lambda}=\frac{\Delta_c}{s\alpha}$ represents the exact spectral resolution given by the SSCSI.

\subsubsection*{$\frac{\Delta_c}{1-s} > \Delta_d$}
\begin{figure}[h!]
	\centering
	\includegraphics[scale=0.35]{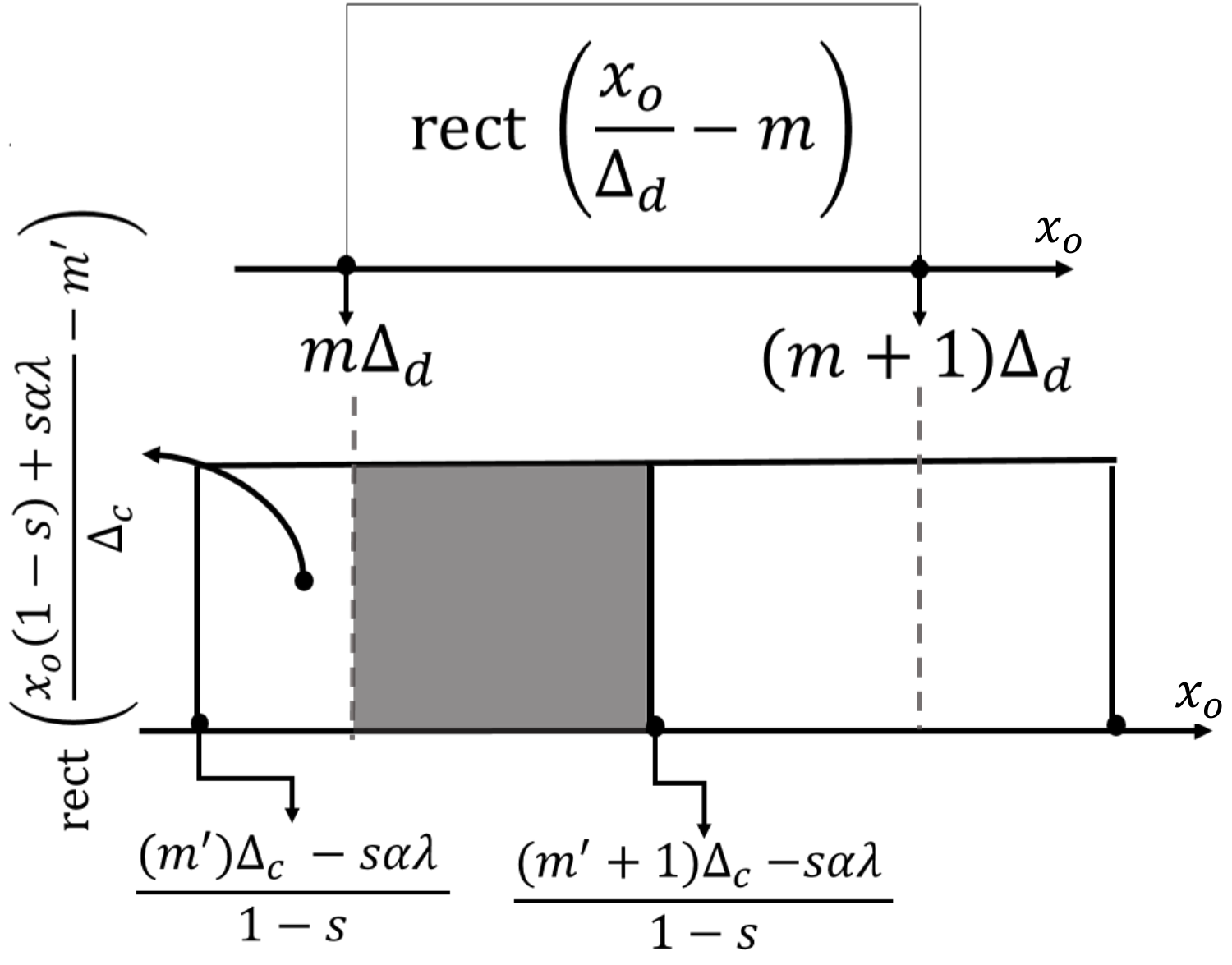}
	\caption{Graphical explanation to calculate the spectral resolution of the SSCSI, when $\frac{\Delta_c}{1-s} > \Delta_d$.}
	\label{ANNEX_B2}
\end{figure}

A similar analysis to the previous one can be done to find the spectral resolution in this case. Figure \ref{ANNEX_B2} shows the overlapping of the two rectangular functions given in Eq. (\ref{spec_2}). Again two scenarios must be analyzed, In the first one (gray area in Fig. \ref{ANNEX_B2}),  $ \frac{(m'+1)\Delta_c-s\alpha\lambda}{1-s} \leq (m+1)\Delta_d$ and $\frac{(m'+1)\Delta_c-s\alpha\lambda}{1-s} \geq (m)\Delta_d$, or $\frac{(m'+1)\Delta_c-(m+1)\Delta_d(1-s)}{s\alpha} \leq \lambda \leq \\ \frac{(m'+1)\Delta_c-(m)\Delta_d(1-s)}{s\alpha}$,  which has an extension of $\Delta_{\lambda}=\frac{\Delta_d(1-s)}{s\alpha}$. In the second case (gray area in Fig. \ref{ANNEX_B2} occupies the whole sensor element), $\frac{(m')\Delta_c-s\alpha\lambda}{1-s} \leq m\Delta_d$ and
$\frac{(m'+1)\Delta_c-s\alpha\lambda}{1-s} \geq (m+1)\Delta_d$, or equivalently
$\frac{m'\Delta_c-m\Delta_d(1-s)}{s\alpha} \leq \lambda \leq \frac{(m'+1)\Delta_c-(m+1)\Delta_d(1-s)}{s\alpha}$,
which has an extension of $\Delta_{\lambda}=\frac{\Delta_c-\Delta_d(1-s)}{s\alpha}$. Notice that, since $\frac{\Delta_c}{1-s}>\Delta_d$, $\frac{\Delta_c}{s\alpha}>\frac{\Delta_d(1-s)}{s\alpha}$ and $\frac{\Delta_c}{s\alpha}>\frac{\Delta_c-\Delta_d(1-s)}{s\alpha}$. Therefore,  $\frac{\Delta_c}{s\alpha}$ can be taken as an upper bound of the spectral resolution.

\vspace{-5mm}
\section*{Appendix C\\ Actual coded aperture shearing at the sensor}
\begin{figure}[h!]
	\centering
	\includegraphics[scale=0.14]{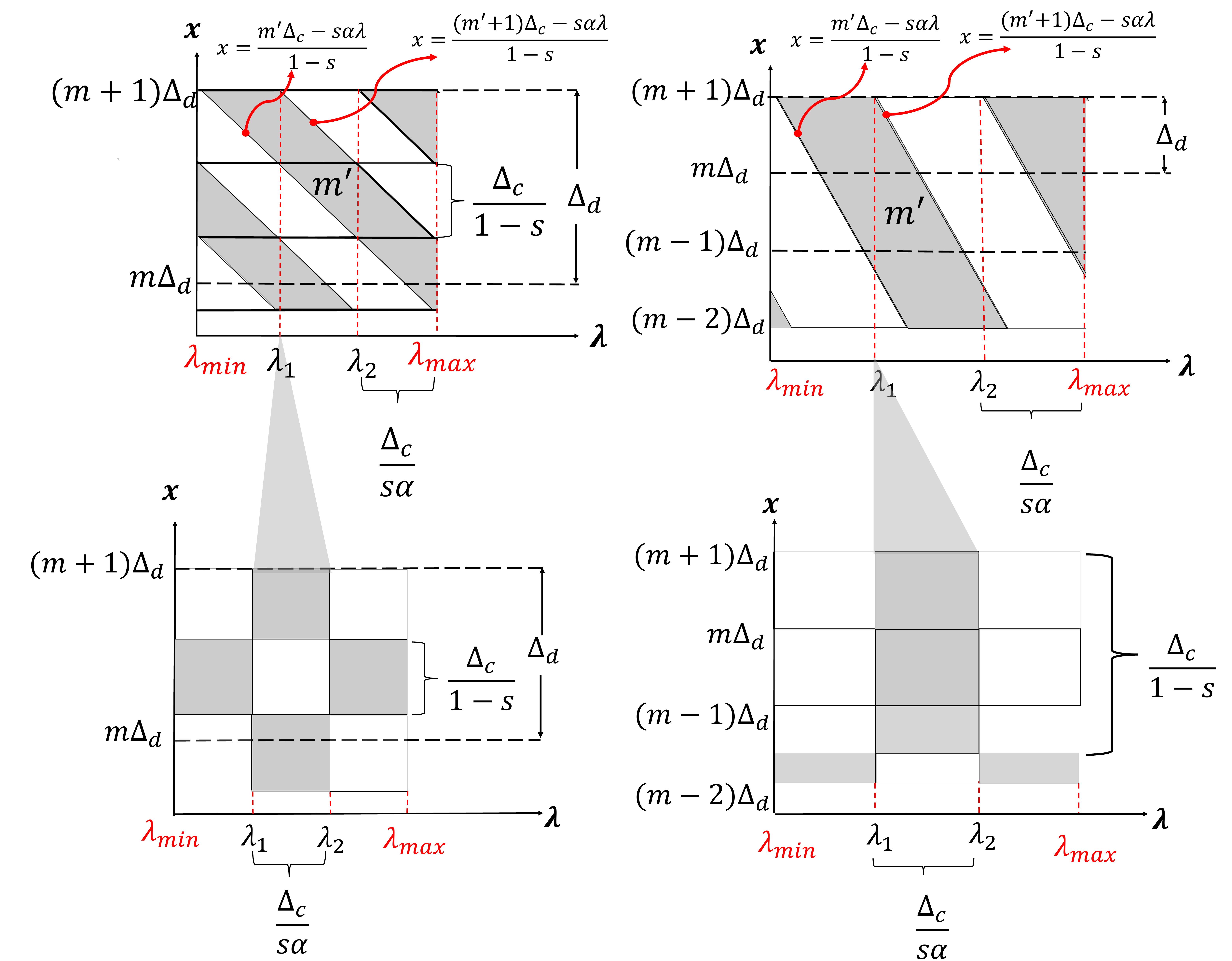}
	\caption{ Top left: Original sheared coded aperture when  $\Delta_c/(1-s) \leq \Delta_d$. Top right: Original sheared coded aperture when $\Delta_c/(1-s) > \Delta_d$. Bottom left: First order approximation when  $\Delta_c/(1-s) \leq \Delta_d$. Bottom right: First order approximation when $\Delta_c/(1-s) > \Delta_d$. Notice how the spectral band $[\lambda_{1},\lambda_{2}]$ by the same pattern that codes the hyperspectral scene at $\lambda_{1}$.}
	\label{discret}
\end{figure}

 Let recall Eq. (\ref{disc_3}), which characterizes the SSCSI sensing process. By analyzing the $x$ dimension of the coded aperture rectangular function in this expression, given by $\mathrm{rect}\left(\frac{x(1-s)+s\alpha\lambda}{\Delta_c}-m',\frac{y}{\Delta_c}-n'\right)$,  one can obtain a figure that depicts the actual sheared pattern. For a given $m'$, $x$ is limited by $\frac{m'\Delta_c-s\alpha\lambda}{1-s}\leq x\leq\frac{(m'+1)\Delta_c-s\alpha\lambda}{1-s}$. Figure \ref{discret}(a) shows the sheared coded aperture seen at the sensor for both cases, when $\Delta_c/(1-s) \leq \Delta_d$ and $\Delta_c/(1-s) > \Delta_d$, in the $x-\lambda$ plane. Notice how $m'$ is bounded by the limits given above. The dashed red lines in this figure represent the spectral division every $\frac{\Delta_c}{s\alpha}$, while the dashed black lines indicate the limits of the $m^{th}$ sensor element. The first order approximation, given by Eqs. (\ref{tot_7}) and (\ref{disc_10}), can be seen in Fig. \ref{discret}(b). Notice how, for example, the spectral band $[\lambda_{1},\lambda_{2}]$ is coded by the same pattern that codes the hyperspectral scene at $\lambda_1$. 

\section*{Appendix D\\ Derivation of Eq. (\ref{s_aprox})}

The number of resolvable bands in the SSCSI system is given by $L=\frac{s\alpha(\lambda_{max}-\lambda_{min})}{\Delta_c}$, where $\alpha(\lambda_{max}-\lambda_{min})$ is the width of the spectral plane. Let $\beta=\alpha(\lambda_{max}-\lambda_{min})/N_c\Delta_c$, where $N_c\Delta_c$ is the coded aperture width. The number of resolvable bands as a function of $\beta$ can be rewritten as $L=s\beta N_c$. If it is assumed that $N_d\Delta_d=N_c\Delta_c$, which means that the coded aperture width is the same as that of the sensor width, last expression can be written as Eq. (\ref{s_aprox}). \\
The ceiling operator $\lceil \cdot \rceil$ in Eqs. (\ref{bands}) and (\ref{s_aprox}) was introduced in order fully use the spectral resolution given by the SSCSI.

\section*{Appendix E\\ Discretization Model for $\Delta_c>\Delta_d$}
This paper proposes a SSCSI discretization measurement model for the specific case of $\Delta_d/\Delta_c=C$, where $C$ is an integer greater than or equal to 1. If $\Delta_c=C_2\Delta_d$, where $C_2$ is an integer greater than 1, the condition $\Delta_c/(1-s)>\Delta_d$ holds for any value of $s$ and Eq. (\ref{disc_10}) can be rewritten as follows:
\begin{multline}
\mathbf{g}_{m,n}=\sum_{k=0}^{L-1}\left(\mathbf{t}_{m'+k-1,n''}\times p_m +\right. \\ 
\left. \mathbf{t}_{m'+k,n''}\times(1-p_m)\right) \mathbf{f}_{m,n,k},
\label{disc_11} 
\end{multline}
where $m'=\left \lfloor \frac{(m)\Delta_d(1-s)+s\alpha\lambda_{min}}{\Delta_c} \right \rfloor +1$, $p_m$ is defined in Eq. (\ref{p_2}) (for $\lambda_o=\lambda_{min}$), and $n''= \left \lfloor \frac{n}{C_2} \right \rfloor$. Notice that the minimum spatially resolvable element in the reconstructed scene is given by $\Delta_d \times \Delta_d$, and the spatial size of the recovered datacube is equal to $N_d\times N_d$.
\medskip
\bibliographystyle{IEEEtran}
\bibliography{bibliography_final}
\end{document}